\documentclass[12pt]{article}
\usepackage{graphicx,psfrag,epsf}
\usepackage{setspace}
\usepackage{natbib}
\usepackage{epsfig}
\usepackage[centertags]{amsmath}
\usepackage{amsfonts}
\usepackage{bbm}     
\usepackage{dsfont}  
\usepackage[figuresright]{rotating}

\newcommand{\ds}{\displaystyle}

\def\R{\mathds{R}}
\def\N{\mathds{N}}
\newcommand\ind{\mathds{1}}

\def\z{{\bf z}}
\def\Z{{\bf Z}}

\def\argmax{\mathop{\rm arg\,max}\limits}

\newtheorem{theorem}{Theorem} 
\newtheorem{lemma}{Lemma} 
\newtheorem{remark}{Remark} 
\newcommand{\indicator}[1]{\mathds{1}_{\left[ {#1} \right] }}

\title{Quantile-based classifiers}
\author{Christian Hennig and Cinzia Viroli
}

\usepackage{setspace}
\date{}
\begin{document}
\maketitle


\begin{abstract}
Quantile classifiers for potentially high-dimensional data are defined by classifying an observation according to a sum of appropriately weighted component-wise distances of the components of the observation to the within-class quantiles. An optimal percentage for the quantiles can be chosen by minimizing the misclassification error in the training sample.

It is shown that this is consistent, for $n \to \infty$, for the classification rule with asymptotically optimal quantile, and that, under some assumptions, for $p\to\infty$ the probability of correct classification converges to one. The role of skewness of the involved variables is discussed, which leads to an improved classifier.

The optimal quantile classifier performs very well in a comprehensive simulation study and a real data set from chemistry (classification of bioaerosols) compared to nine other classifiers, including the support vector machine and the recently proposed median-based classifier (\cite{HTX09}), which inspired the quantile classifier.

\end{abstract}


\section{Introduction}
Supervised classification is a major issue in statistics and has received a wide interest in the scientific literature of many disciplines.

The ``large microcosm'' of classification methods \citep{HAN97} can be broadly divided into parametric methods, which make distributional assumptions about the data, and nonparametric methods, which alternatively concentrate on the local vicinity of the point to be classified, such as nearest neighbor methods \citep{CH67} and kernel smoothing \citep{MRW99}.

Parametric methods use the estimated class conditional distributions for the construction of the classification rule. The traditional linear and quadratic discriminant analysis, mixture discriminant analysis \citep{HT96}, the naive Bayes probabilistic model \citep{JL95,HY01}, model-based discriminant analysis \citep{BC96,FR02} and nonlinear neural networks \citep{RIP94} are examples of such methods. See also \cite{FRI89,GUO07,CTL11} and the references therein. Implementing such methods in high dimensional settings, which are very common nowadays, can be cumbersome and computationally demanding, because of the well-known curse of dimensionality \citep{BEL61}.

A great deal of work, especially on distance-based methods, has been carried out to try to circumvent this problem. Distance-based classifiers only use partial information of the class conditional distributions, typically central moments. Centroid-based methods have been successfully used for gene expression data \citep{THNC02,DUD02,DAB05,FAN08}. Median-based classifiers \citep{Jor04,GHO05} represent a more robust alternative in problems where distributions have heavy tails. \cite{HTX09} proposed a component-wise median based classifier which behaves well in high dimensional space. It assigns a new observed vector to the class having the smallest $L_1$-distance from the class conditional  component-wise median vectors of the training set.

All these methods consider the distance from the ``core'' of a distribution as the major source of the discriminatory information. But tails may be important as well and may contain relevant information. It may therefore be fruitful to go beyond the central moments.

In this work we define and explore a family of classifiers based on the quantiles of the class conditional distributions. The idea was originally inspired by the component-wise median classifier \citep{HTX09}.

More specifically, by using the natural distance for quantiles, we will obtain the component-wise quantile classifier as function of the $\theta$-quantile,
$\theta \in [0,1]$. The optimal $\theta$ chosen in the training set will define the empirically optimal quantile classifier. We will prove the consistency of this choice for the $\theta$ that yields the optimal true correct classification probability as $n \rightarrow \infty$. We will also show under certain assumptions that the correct classification probability converges to one as $p \rightarrow \infty$ together with the sample size, similarly to what \cite{HTX09} did for the component-wise median classifier.

 The paper is organized as follows. In Section 2 we review the distance-based classifiers and define the proposed quantile classifier. The theoretical properties of the method are explored in Sections 3 and 4. A large simulation study and a real application are presented in Section 5.

\section{The classification rule}
\subsection{Distance-based classifiers}
We consider the problem of constructing a quantile distance-based discriminant rule for classifying new observations into one of $g$ populations or classes. Without loss of generality we discuss the problem for $g=2$. Generalization for $g>2$ is straightforward.

Let $\Pi_0$ and $\Pi_1$ be two populations with probability densities $P_0$ and $P_1$ on $\mathcal{R}^p$.
Distance based classifiers \citep{Jor04,THNC03,HTX09} assign a new data value $\textbf{z}=(z_1,\ldots,z_p)$ to the population from which it has lowest distance. More specifically, the decision rule allocates $\textbf{z}$ to $\Pi_0$ if
\begin{eqnarray}\label{e:decision}
\sum_{j=1}^p \{d({z}_j,{Y}_j)-d({z}_j,{X}_j)\}>0,
\end{eqnarray}
where $\textbf{X}=\{X_1,\ldots,X_{p}\}$ and $\textbf{Y}=\{Y_1,\ldots,Y_{p}\}$ are $p$-variate random variables from populations $\Pi_0$ and $\Pi_1$ and $d(\cdot)$ denotes a specific distance measure. Expression (\ref{e:decision}) represents a rather general discriminant rule formulation that includes centroid classifiers \citep{THNC02,THNC03,WZ07}, the recent component-wise median-based classifiers \citep{HTX09}, and other variants by differently specifying the distance measure $d(\cdot)$. On the other hand, summing up component-wise differences means that correlation between variables is not taken into account. If $p$ is small and there are many observations, this is rather restrictive. However, if $p$ is large and the number of observations is rather low, it can be effective to avoid overfitting.
By considering the Euclidean distance between $z_j$ and the expectations of $X_j$ and $Y_j$, the component-wise centroid classifier assigns $\textbf{z}$ to $\Pi_0$ if
\begin{eqnarray}\label{e:centroid}
\sum_{j=1}^p \{({z}_j-E({Y}_j))^2-({z}_j-E({X}_j))^2\}>0,
\end{eqnarray}
and to $\Pi_1$ otherwise.
By taking the $L_1$ (Manhattan)-distance between $z_j$ and the medians of $X_j$ and $Y_j$ the component-wise median-based classification rule can be defined as
\begin{eqnarray}\label{e:median}
\sum_{j=1}^p \{|{z}_j-\textrm{med} ({Y}_j)|-|{z}_j-\textrm{med}({X}_j)|\}>0.
\end{eqnarray}
Note that in realistic situations neither $P_0$ and $P_1$ nor their moments
are known. We rather observe two sets $\textbf{x}_1,\ldots,\textbf{x}_{n_0}$ and $\textbf{y}_1,\ldots,\textbf{y}_{n_1}$ from $\Pi_0$ and $\Pi_1$; they represent the training data samples from which the desirable moments must be inferred. For instance, the sample version of the centroid classifier assigns $\textbf{z}$ to $\Pi_0$ if
\begin{eqnarray}\label{e:decision2}
\sum_{j=1}^p \{({z}_j - {\bar{y}}_j)^2-({z}_j - \bar{x}_j)^2\}>0,
\end{eqnarray}
where $\bar{y}_j$ and  $\bar{x}_j$ denote the $j$th component of the sample mean vectors. Analogously, the sample version of the discriminant rule (\ref{e:median}) requires computing the empirical component-wise medians. \cite{HTX09} stated that median classifiers are more robust against heavy tails of the data distribution than centroid classifiers, thanks to the metric $L_1$ instead of $L_2$, and they provided a formal proof of the fact that asymptotically the correct decision is made by the rule with probability one, if
the dimension as well as the numbers of observations in both classes tend
to infinity under some further assumptions.

The choice of the metric $L_1$, instead of $L_2$, in the median classifier addresses the need of consistency between metric and related minimizer moment; in fact, the mean vector (centroid) is the statistic that minimizes the sum of $L_2$-distances of points to the centroid, whereas the median minimizes the sum of the corresponding $L_1$-distances. Hybrid alternatives may exist, such as an $L_1$-version of the centroid classifier. However, they look convincing from neither a theoretical nor a practical point of view. Not only does a hybrid alternative mismatch the relation between metric and related minimizer quantity, but it also seems to produce higher misclassification rates in practice (see, for instance, \cite{HTX09}).

\subsection{The quantile classifier}

We introduce the family of the component-wise quantile classifiers that includes the median classifier as special case.
By definition, the $\theta^{th}$ quantile of a univariate random variable $X$ with probability distribution function $F_X$, denoted by $q_X(\theta)$, is the solution to $q_X(\theta)=F_X^{-1}(\theta)=\inf\{x:F_X(x)\geq\theta\}$, with $\theta \in [0,1]$.
Analogously to the roles of median and centroid with respect to the $L_1$- and $L_2$-metric, the $\theta^{th}$ quantile of $F_X$ is the value $q$ that minimizes the following population distance
\begin{eqnarray}\label{e:distance}
\theta\int_{x>q}|x-q|dF_X(x)+(1-\theta)\int_{x<q}|x-q|dF_X(x).
\end{eqnarray}
This can be easily proven by observing that (\ref{e:distance}) is minimized for $F_X(q)=\theta$.
Given a set of observations $x_1,x_2,\ldots,x_n$, the empirical $\theta^{th}$ quantile of $X$ can be found by minimizing the sample counterpart of (\ref{e:distance}):

\begin{eqnarray}\label{e:distance2}
\theta \sum_{x_i>q}|x_i-q|+(1-\theta)\sum_{x_i \leq q}|x_i-q|
= \sum_{x_i} \left(\theta+(1-2\theta)\indicator{x_i \leq q}\right)|x_i-q|.
\end{eqnarray}

 The metric (\ref{e:distance2}) is used to define the component-wise quantile-based new classifier. Given two sets of observations from the two populations $\Pi_0$ and $\Pi_1$, $\textbf{x}_1,\ldots,\textbf{x}_{n_0}$ and $\textbf{y}_1,\ldots,\textbf{y}_{n_1}$, a new observation $\z=(z_1,\ldots,z_p)\in\R^p$ is assigned to $\Pi_0$ if
\begin{eqnarray}\label{e:quantile}
\sum_{j=1}^p \left[
\left(\theta+(1-2\theta)\indicator{z_j \leq q_{1j}(\theta)}\right)|z_j-q_{1j}(\theta)| -
\left(\theta+(1-2\theta)\indicator{z_j \leq q_{0j}(\theta)}\right)|z_j-q_{0j}(\theta)|\right] >0,
\end{eqnarray}
where $q_{0j}(\theta)$ and $q_{1j}(\theta)$ are the marginal quantile functions of the two class-distributions evaluated at a fixed value of $\theta$.

For $j=1,\ldots,p$ and $k=0,1$,
let $\Phi_{j}(\z,\theta,q)=\left(\theta+(1-2\theta)\indicator{z_j\le q}\right)|z_j-q|$ and
$\Phi_{kj}(\z,\theta)=\left(\theta+(1-2\theta)\indicator{z_j\le q_{kj}(\theta)}\right)|z_j-q_{kj}(\theta)|$. Then, for fixed $\theta$, the classification rule (\ref{e:quantile}) is equivalent to assigning $\textbf{z}$ to $\Pi_0$ if $\sum_{j=1}^p \Phi_{0j}(\z,\theta)<\sum_{j=1}^p \Phi_{1j}(\z,\theta)$, and to $\Pi_1$ otherwise.

\begin{remark}
The applicability of the decision rule (\ref{e:quantile}) to more than $g=2$ classes is straightforward. By definition, the quantile classifier rule for allocating an observation $\z$ to one of $g$ populations $\Pi_1,\ldots,\Pi_g$ is to allocate $\z$ to the population which gives the lowest quantile distance $\sum_{j=1}^p \Phi_{kj}(\z,\theta)$, with $k=1,\ldots,g$.
\end{remark}

\begin{remark}
Note that for $\theta=0.5$ the objective function in (\ref{e:distance2}) (multiplied by 2) is the $L_1$-distance between $x$ and the median. Therefore decision rule (\ref{e:quantile}) coincides with the component-wise median classifier when $\theta=0.5$.
\end{remark}

Given the two populations, $\Pi_0$ and $\Pi_1$ with prior probabilities $\pi_0$ and $\pi_1$, respectively, the probability of correct classification of the quantile classifier is
\begin{eqnarray}\label{e:probcc}
  \Psi(\theta) &=& \pi_0\int \indicator{\ds\sum_{j=1}^p (\Phi_{1j}(\z,\theta)-
\Phi_{0j}(\z,\theta))>0} dP_0(\z)+ \nonumber \\
&& \pi_1\int \indicator{\ds\sum_{j=1}^p (\Phi_{1j}(\z,\theta)-
\Phi_{0j}(\z,\theta))\le 0} dP_1(\z).
\end{eqnarray}

This quantity represents the theoretical rate of correct classification based on the true quantiles. This rate can be used to measure the performance of the discriminant rule with respect to the chosen value $\theta$ regardless of the sample size (we will later simulate such rates based on empirical quantiles, as relevant in real applications). The following lemma provides a useful formula to derive the theoretical rate of correct classification as function of $\theta$ for $p=1$.

\begin{lemma}\label{l:probcorrect} When $p=1$, the probability of correct classification of the quantile classifier takes the following simple form.
\begin{description}
  \item[-] If $q_0(\theta) \leq q_1(\theta)$,
  \begin{eqnarray}\label{e:theo1a}
  \Psi(\theta) = \pi_0F_0(\ddot{\theta})+\pi_1(1-F_1(\ddot{\theta}))
  \end{eqnarray}
   with $\ddot{\theta}=\theta q_0(\theta)+(1-\theta)q_1(\theta)$.
  \item[-] If $q_0(\theta) > q_1(\theta)$,
  \begin{eqnarray}\label{e:theo1b}
  \Psi(\theta) = \pi_1F_1(\dot{\theta})+\pi_0(1-F_0(\dot{\theta}))
  \end{eqnarray}
  with $\dot{\theta}=\theta q_1(\theta)+(1-\theta)q_0(\theta)$.
\end{description}
where $q_0(\theta)$ and $q_1(\theta)$ are the true quantiles of the two populations.
\end{lemma}

{\bf Proof} of Lemma \ref{l:probcorrect}.

Consider that in the univariate case $\Phi_{0}(z,\theta)$ and $\Phi_{1}(z,\theta)$ may be rewritten as
\begin{eqnarray*}
\Phi_{0}(z,\theta)=(1-\theta)\left(q_0(\theta)-z\right)\indicator{z \leq q_0(\theta)} + \theta\left(z-q_0(\theta)\right)\indicator{z > q_0(\theta)} \\
\Phi_{1}(z,\theta)=(1-\theta)\left(q_1(\theta)-z\right)\indicator{z \leq q_1(\theta)} + \theta\left(z-q_1(\theta)\right)\indicator{z > q_1(\theta)}
\end{eqnarray*}

For a fixed $\theta$, the integral (\ref{e:probcc}) can be easily solved by splitting it into four parts according to the possible disjoint regions of the domain of $Z$ with respect to $q_0(\theta)$ and $q_1(\theta)$, namely: (a) $z \leq \min(q_0(\theta),q_1(\theta))$, (b) $q_0(\theta) < z \leq q_1(\theta)$, (c) $q_1(\theta) \leq z \leq q_0(\theta)$ and (d) $z > \max(q_0(\theta),q_1(\theta))$.

If $z \leq \min(q_0(\theta),q_1(\theta))$ the integral becomes
\begin{eqnarray*}
  \Psi_a(\theta) &=& \pi_0\int_{-\infty}^{\min(q_0(\theta),q_1(\theta))} \indicator{(1-\theta)(q_1(\theta)-q_0(\theta))>0} dP_0(z) \nonumber \\ &+& \pi_1\int_{-\infty}^{\min(q_0(\theta),q_1(\theta))} \indicator{ (1-\theta)(q_1(\theta)-q_0(\theta))\le 0} dP_1(z) \nonumber \\
  &=& \pi_0\int_{-\infty}^{q_0(\theta)} dP_0(z) \indicator{q_1(\theta)>q_0(\theta)}  + \pi_1\int_{-\infty}^{q_1(\theta)} dP_1(z) \indicator{q_1(\theta) \leq q_0(\theta)} \nonumber \\ &=&\pi_0\theta \indicator{q_1(\theta)>q_0(\theta)}  + \pi_1\theta \indicator{q_1(\theta) \leq q_0(\theta)}.
\end{eqnarray*}
In the second case the integral is
\begin{eqnarray*}
  \Psi_b(\theta) &=& \pi_0\int_{q_0(\theta)}^{q_1(\theta)} \indicator{(1-\theta) (q_1(\theta)-z) - \theta(z-q_0(\theta)) >0} dP_0(z) \nonumber \\ &+& \pi_1\int_{q_0(\theta)}^{q_1(\theta)} \indicator{(1-\theta) (q_1(\theta)-z) - \theta(z-q_0(\theta))\le 0} dP_1(z) \nonumber \\
  &=& \pi_0\int_{q_0(\theta)}^{\theta q_0(\theta)+(1-\theta)q_1(\theta)} dP_0(z)\indicator{q_0(\theta) \leq q_1(\theta)} \nonumber \\ &+& \pi_1\int_{\theta q_0(\theta)+(1-\theta)q_1(\theta)}^{q_1(\theta)} dP_1(z) \indicator{q_0(\theta) \leq q_1(\theta)}
  .
\end{eqnarray*}
Similarly, for the cases (c) and (d) the integrals are
\begin{eqnarray*}
  \Psi_c(\theta) &=& \pi_0\int_{\theta q_1(\theta)+(1-\theta)q_0(\theta)}^{q_0(\theta)} dP_0(z)\indicator{q_1(\theta) \leq q_0(\theta)} \nonumber \\ &+& \pi_1\int_{q_1(\theta)}^{\theta q_1(\theta)+(1-\theta)q_0(\theta)} dP_1(z) \indicator{q_1(\theta) \leq q_0(\theta)},
\end{eqnarray*}
and
\begin{eqnarray*}
  \Psi_d(\theta) &=& \pi_0(1-\theta) \indicator{q_0(\theta)>q_1(\theta)}  + \pi_1(1-\theta) \indicator{q_0(\theta) \leq q_1(\theta)}
  .
\end{eqnarray*}
Now, when $q_0(\theta) \leq q_1(\theta)$, $\Psi(\theta)$ is the sum of $\Psi_a(\theta)$, $\Psi_b(\theta)$ and $\Psi_d(\theta)$ corresponding to disjoint domain regions of $Z$:
\begin{eqnarray*}
  \Psi(\theta) &=& \pi_0 \theta +\pi_0\int_{q_0(\theta)}^{\theta q_0(\theta)+(1-\theta)q_1(\theta)} dP_0(z)+ \pi_1\int_{\theta q_0(\theta)+(1-\theta)q_1(\theta)}^{q_1(\theta)} dP_1(z)  + \pi_1(1-\theta) \nonumber \\
  &=& \pi_0 \theta +\pi_0F_0(\theta q_0(\theta)+(1-\theta)q_1(\theta)) - \pi_0 \theta + \pi_1\theta \nonumber \\ && -  \pi_1 F_1(\theta q_0(\theta)+(1-\theta)q_1(\theta))  + \pi_1(1-\theta) \\ &=&\pi_0F_0(\ddot{\theta})+\pi_1(1-F_1(\ddot{\theta})).
\end{eqnarray*}
Analogously, when $q_0(\theta) > q_1(\theta)$, $\Psi(\theta)$ is the sum of $\Psi_a(\theta)$, $\Psi_c(\theta)$ and $\Psi_d(\theta)$ from which:
\begin{eqnarray*}
  \Psi(\theta) &=& \pi_1 \theta + \pi_0\int_{\theta q_1(\theta)+(1-\theta)q_0(\theta)}^{q_0(\theta)} dP_0(z)+ \pi_1\int_{q_1(\theta)}^{\theta q_1(\theta)+(1-\theta)q_0(\theta)} dP_1(z) + \pi_0(1-\theta)
  \nonumber \\ &=& \pi_1F_1(\dot{\theta})+\pi_0(1-F_0(\dot{\theta})).
\end{eqnarray*}

\begin{figure}
  \includegraphics[scale=0.8]{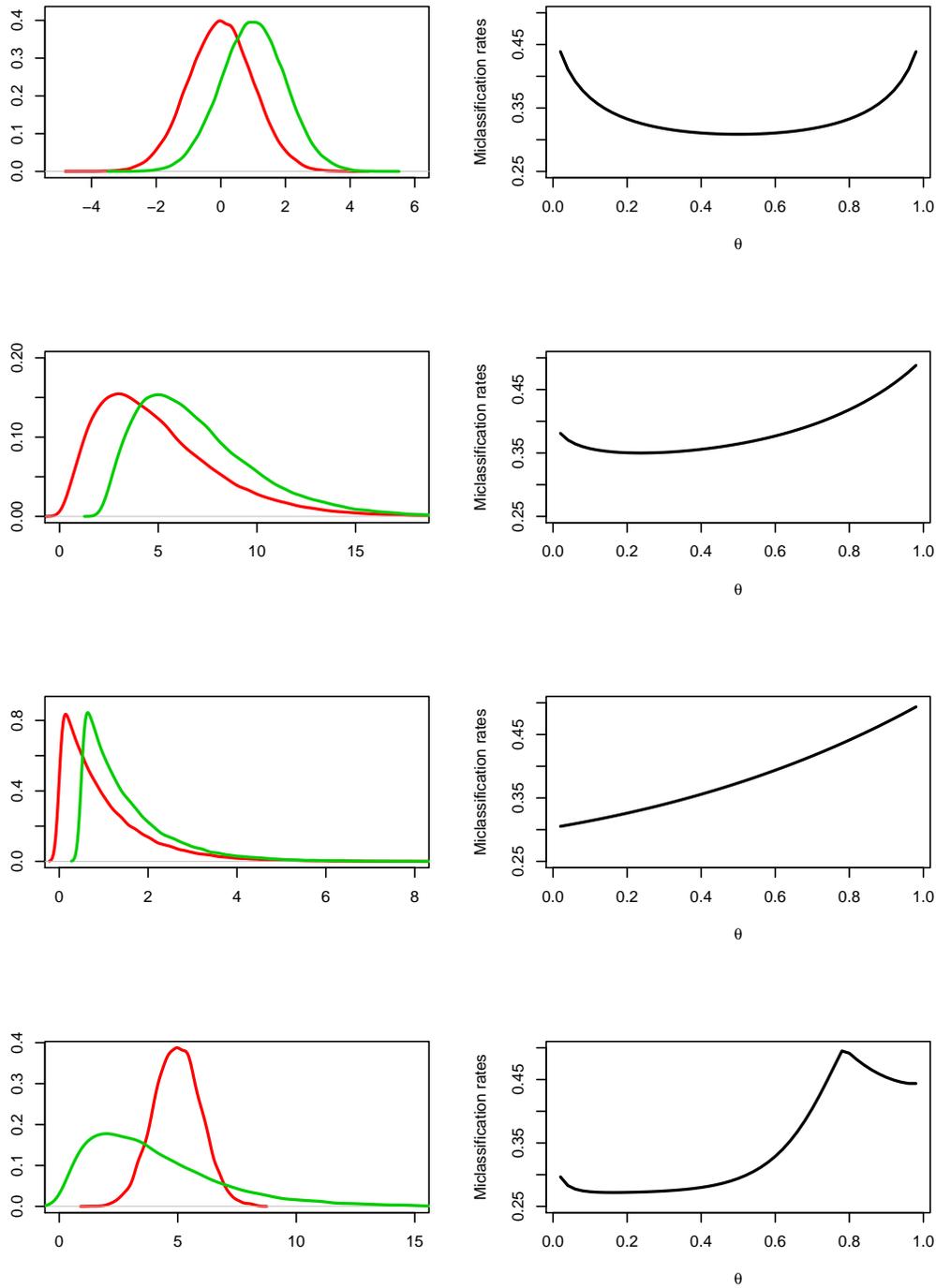}
  \caption{Theoretical misclassification rates $1-\Psi(\theta)$ for four different scenarios. First row: probability density functions of two location-shifted Gaussians and corresponding misclassification function of $\theta$. Second row: two location-shifted chi-squared distributions. Third row: two location-shifted exponentials. Last row: a Gaussian vs a chi-squared distribution.}
  \label{f:theomiscrates}
\end{figure}

Lemma \ref{l:probcorrect} provides a direct formula to compute the probability of correct classification - analytically or numerically - for given values of $\theta$.
Suppose the two populations $\Pi_0$ and $\Pi_1$ have exponential distributions but differ for a location shift $c$: $X \sim P_0=Exp(\lambda)$ and $Y \sim P_1=Exp(\lambda)+c$, with $c>0$. Then $F_0(x)=1-\exp(-\lambda x)$ and $F_1(y)=1-\exp(-\lambda (y-c))$. Since the probability distribution functions of the exponentials can be expressed in closed form, the two quantile functions can be analytically derived by solving $F_0(x)^{-1}$ and $F_1(y)^{-1}$, from which $q_0(\theta)=-\frac{\ln(1-\theta)}{\lambda}$ and $q_1(\theta)=-\frac{\ln(1-\theta)}{\lambda}+c$, respectively.
Since $c>0$, we have $q_0(\theta) \leq q_1(\theta) \ \forall \theta \in [0,1]$. By applying (\ref{e:theo1a}), we get the rates of correct classification of the quantile classifier for two (varying-location) exponential distributions as a function of $\theta$:
$$\Psi(\theta) = \pi_0-(1-\theta)e^{c \lambda\theta}(\pi_0 e^{-c\lambda}-\pi_1).$$

 Figure \ref{f:theomiscrates} (second panel of third row) shows the theoretical misclassification rates, $1-\Psi(\theta)$, of two exponential populations with $\lambda=1$, $c=0.5$ and $\pi_0=\pi_1=0.5$. It is interesting to note that the minimum misclassification rate can be obtained for $\theta$ approaching zero.
 This particular choice for $\theta$ is related to the high level of skewness of the exponential distribution. To make this clearer, we also considered further scenarios, namely two location-shifted Gaussians, $\mathcal{N}(0,1)$ and $\mathcal{N}(1,1)$, and two location-shifted chi-squared distributions with 5 degrees of freedom and shift $c=2$ (first and second rows of Figure \ref{f:theomiscrates}). The theoretical misclassification rates, $1-\Psi(\theta)$, can be easily obtained numerically. In the Gaussian scenario the minimum value of $1-\Psi(\theta)$ is obtained for $\theta=0.5$. This is not surprising because of the symmetric shape of the Gaussian. But more asymmetric distributions (second and third rows in Figure \ref{f:theomiscrates}) tend to yield an optimum $\theta$ far away from the midpoint 0.5, with positive skewness normally associated with the optimum being below 0.5 and negative skewness with an optimum above 0.5 (obviously, if skewness is reversed by multiplying a random variable by -1, the resulting optimal $\theta$ will be one minus the original optimum). This indicates that the best $\theta$ for one problem is not the best for another, and this choice is of crucial importance. For example, in the second case, the theoretical quantile function is minimized for $\theta=0.236$. The fourth row of Figure \ref{f:theomiscrates} shows the classification problem with two differently distributed populations, a Gaussian distribution with parameters 5 and 1 and a chi-squared distribution with 4 degrees of freedom. The optimal quantile classifier corresponds to $\theta=0.162$.

\begin{figure}
  \includegraphics[scale=0.8]{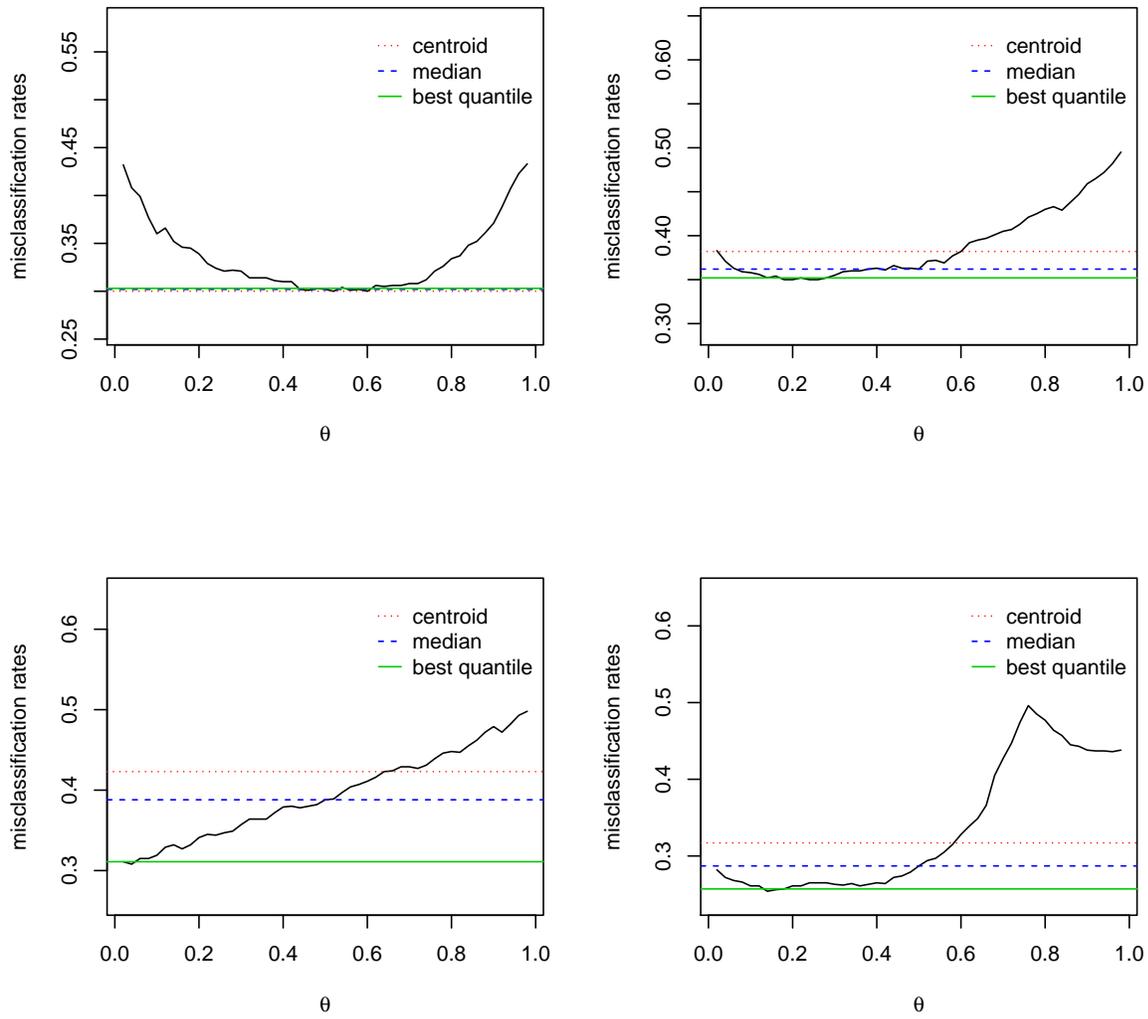}
  \caption{Misclassification rates obtained in the test set of a simulation study. For comparative purposes, the horizontal lines indicate the misclassification rates of the median classifier, of the centroid classifier and of the optimal quantile classifier in the training set.}
  \label{f:estmiscrates}
\end{figure}

 Figure \ref{f:estmiscrates} shows the estimated misclassification rates obtained in the four scenarios by a simulation study with sample sizes of training set and test set equal to 500. The plotted line is the empirical curve of the misclassification rate obtained in the test set for different values of $\theta$. It approximates the theoretical one well. The horizontal lines indicate the misclassification rates obtained by the centroid classifier, the median classifier and quantile classifier corresponding to the optimal value $\theta$ chosen in the training set.

Unfortunately, Lemma \ref{l:probcorrect} cannot easily be extended to the multivariate setting, unless some very restrictive conditions are assumed regarding independence of the variables and strict rules about the ranking of the $p$ different quantiles $q_{kj}(\theta) \ j=1,\ldots,p$ within each population.

\subsection{The empirically optimal quantile classifier}
\label{sempopt}
In real applications the problem of the choice of the quantile value in the family of possible quantile classifiers can be addressed by selecting the
optimum $\theta$ based on misclassification rates in the training sample.
This leads to the definition of the empirically optimal quantile classifier.

First, we introduce some notation.
Let $(\Z_1,C_1),(\Z_2,C_2),\ldots$ be i.i.d. $\R^p\times \{0,1\}$-valued RV.
Let $\Z_1$ be distributed according to a 2-component mixture of distributions
$P_0={\cal L}(\Z_1|C_1=0)$ and $P_1={\cal L}(\Z_1|C_1=1)$. Let
$\pi_0=P\{C_1=0\},\ \pi_1=1-\pi_0$. Let
$P_{01},\ldots,P_{0p}$ denote the marginal distributions of $P_0$, analogously
$P_{11},\ldots,P_{1p}$.

For arbitrarily small $0<\tau<\frac{1}{2}$ define
$T=[\tau,1-\tau]$.
For $\theta\in [0,1],\ j=1,\ldots,p,\ k=0,1$ denote
$q_{kj}(\theta)$ the $\theta$-quantile of $P_{kj}$.
For given $(\Z_1,C_1),\ldots,(\Z_n,C_n)$ let $q_{kjn}(\theta)$ be the empirical
$\theta$-quantile for the subsample defined by $C_i=k,\ i=1,\ldots,n$.

For $j=1,\ldots,p,\ k=0,1,\ \z=(z_1,\ldots,z_p)\in\R^p$,
let
$\Phi_{j}(\z,\theta,q)=\left(\theta+(1-2\theta)\ind[z_j\le q]\right)|z_j-q|$
(in abuse of notation, assumption B2 of Theorem \ref{tpinfty} will apply
$\Phi_j$ to infinite-dimensional $\z$).
$\Phi_{kj}(\z,\theta)$ is used for
$\Phi_{j}(\z,\theta,q_{kj}(\theta))$. $\Phi_{kjn}(\z,\theta)$ is used for
$\Phi_{j}(\z,\theta,q_{kjn}(\theta))$.

The empirically optimal quantile classifier is
defined by assigning $\textbf{Z}$ to $\Pi_0$ if
\begin{equation}
  \label{eq:qclass}
  \sum_{j=1}^p (\Phi_{1jn}(\Z,\theta_n)-\Phi_{0jn}(\Z,\theta_n))>0,
\end{equation}
where $\theta_n=\argmax_{\theta\in T}\Psi_n(\theta)$ is the estimated optimal
$\theta$ from $(\Z_1,C_1),\ldots,(\Z_n,C_n)$ (if the argmax is not unique,
any maximizer can be chosen), and
the observed rate of correct classification in data $(\z_1,c_1),\ldots,(\z_n,c_n)$
is
\begin{eqnarray*}
\Psi_n(\theta) &=& \frac{1}{n}\left(\ds\sum_{i:\ c_i=0} \indicator{\ds\sum_{j=1}^p (\Phi_{1jn}(\z_i,\theta)-
\Phi_{0jn}(\z_i,\theta))>0} +\right. \\
&& \left.\ds\sum_{i:\ c_i=1} \indicator{\ds\sum_{j=1}^p (\Phi_{1jn}(\z_i,\theta)-
\Phi_{0jn}(\z_i,\theta))\le 0}\right).
\end{eqnarray*}
Note that we look for the optimal value of $\theta$ in $T$, a closed interval
not containing zero. In practice, a small nonzero $\tau$ needs to be chosen, and $\Psi_n(\theta)$ is evaluated on a grid of equispaced values between $\tau$
and $1-\tau$. $T$
will in practice depend on the number of observations. $\tau$ should be
chosen small but large enough that there is still a certain amount of
information to estimate the $\tau$-quantile. $\tau$ should not be seen as a
crucial tuning parameter of the method; we recommend to choose it as small as
possible in order to find the empirical optimum of $\theta$, only making sure
that the estimated $\tau$-quantile still is of some use.

In case of a tie (i.e., equal training set misclassification rates for different values of $\theta$, which can easily happen for data sets with small $n$), we recommend to fit a square polynomial to the misclassification rate as function of $\theta$ and to choose the optimum $\theta$ according to this fit out of the empirically optimal ones.

\begin{remark}\label{rstan}
As well as a number of other classifiers, the quantile
classifier depends on the scaling of the variables.
This dependence can be removed by standardizing the variables.
Straightforward ways of doing this would be standardization to
unit variance, range, or interquartile range.
Standardization can be seen as implicit reweighting of the variables.
Optimally, variables are treated in such a way that their relative
weights reflect
their relative information contents for classification.

This means that in practice, in some situations, standardizing is not
advisable, namely where variables have the same measurement units
and there are subject-matter reasons to expect that the information content
of the variables
for classification may be indicated by their variation. Section \ref{sreal}
presents an example for a situation in which the variability of variables
is connected to their information content, and for
a standardization scheme driven by subject knowledge.

Where variables are standardized, standardization to unit pooled
within-class variance (or range, or interquartile range) as estimated from
the training data can be expected to improve matters compared with the
plain variance, because the separation between classes contributes strongly
to the plain variance. This means that variables with a strong
separation between
classes and hence a large amount of classification information will be
implicitly downweighted, whereas standardization to unit pooled within-class
variance will downweight variables for which the classes are heterogeneous and
which are therefore not so useful for classification.

Given enough data, one could
use cross-validation to choose an optimal standardization scheme.
\end{remark}

In the next section, we will present some theoretical properties of the proposed classifier.

\section{Consistency of the quantile classifier} \label{scons}
The asymptotic
probability of correct classification of the quantile classifier is
defined in (\ref{e:probcc}).
Let $\tilde\theta=\argmax_{\theta\in T}\Psi(\theta)$ be the optimal $\theta$
regarding the true model.

The theory needs the following assumptions:
\begin{description}
\item[A1] For all $j=1,\ldots,p,\ k=0,1:\ q_{kj}$ is a continuous function
of $\theta\in T$.
\item[A2] For all $\theta\in T$,
$P\left\{\sum_{j=1}^p (\Phi_{1j}(\Z,\theta)-\Phi_{0j}(\Z,\theta))=0
\right\}=0.$
\end{description}
If A1 and A2 are not
fulfilled, there may be ambiguities regarding the optimal quantile or the
classification of a set of points with nonzero probability. In case of
violation of A2, the problem caused by this will affect a subset of the
data space with at most
the probability given in A2. A1 will probably only affect consistency if
violation happens around the optimal $\theta$, and probably only weakly so if
the discontinuity is small.
 \begin{theorem}\label{tconsist} Assume A1 and A2. Then, for any $\epsilon>0$,
  \begin{displaymath}
    \lim_{n\to\infty}
P\{|\Psi(\tilde\theta)-\Psi(\theta_n)|>\epsilon\}=0.
  \end{displaymath}
\end{theorem}
This means that for $n\to\infty$ the optimal
true correct classification probability equals the true
one corresponding to the empirically optimal $\theta_n$, i.e., the $\theta$
chosen for the quantile classifier, which is therefore asymptotically optimal
(and therefore at least as good as $\frac{1}{2}$, which defines the median
classifier). Theorem \ref{tconsist} is based on $g=2$. This is for convenience of the proof only. Arguments carry over to $g>2$ in a straightforward manner.

{\bf Proof} of Theorem \ref{tconsist}.
\begin{equation}
  \label{eq:aux}
  |\Phi_j(\z,\theta_1,q_1)-\Phi_j(\z,\theta_2,q_2)|\le |z_j|
|\theta_2-\theta_1|+4|q_2-q_1|
\end{equation}
is proved below as Lemma \ref{lineqdiff} for $j=1,\ldots,p,\
\theta_i$-quantiles
$q_i,\ i=1, 2$.
Together with A1, this implies the continuity of $\Psi$, because
for given $\z$, $\Phi_{kj}$ is a continuous function of $\theta$, and
the dominated convergence theorem makes the integrals of the indicator functions
converge for $\theta_n\to\theta$.

The proof of Theorem \ref{tconsist} is now based on
\begin{equation}
  \label{eq:decomp}
|\Psi(\tilde\theta)-\Psi(\theta_n)|\le |\Psi(\tilde\theta)-\Psi_n(\tilde\theta)|
+|\Psi_n(\tilde\theta)-\Psi_n(\theta_n)|+|\Psi_n(\theta_n)-\Psi(\theta_n)|.
\end{equation}
In order to show that all three terms on the right side are asymptotically
small, the following result is proved below as Lemma \ref{lunifpsi}:
\begin{equation}
  \label{eq:sup}
  \forall \epsilon>0:\ \lim_{n\to\infty}
P\left\{\sup_{\theta\in T}|\Psi_n(\theta)-\Psi(\theta)|>\epsilon\right\}=0.
\end{equation}

(\ref{eq:sup}) forces the first and third term on the right side of
(\ref{eq:decomp}) to converge to zero in probability. Consider now
the second term. By definition,
\begin{displaymath}
  \Psi_n(\theta_n)\ge \Psi_n(\tilde\theta),\ \Psi(\tilde\theta)\ge \Psi(\theta_n).
\end{displaymath}
Using (\ref{eq:sup}) again, for large $n$ both
$|\Psi_n(\tilde\theta)-\Psi(\tilde\theta)|$ and
$|\Psi_n(\theta_n)-\Psi(\theta_n)|$ will be arbitrarily small with
arbitrarily large probability, and this
makes $|\Psi_n(\tilde\theta)-\Psi_n(\theta_n)|$ arbitrarily small, too.
Altogether, this proves the theorem.

\begin{lemma}\label{lineqdiff}
(\ref{eq:aux}) holds for
$j\in\{1,\ldots,p\}, \theta_1, \theta_2\in (0,1),\ q_1, q_2\in\R$,
assuming $\theta_1\le\theta_2\Rightarrow q_1\le q_2$ and analogously
for ``$\ge$'' (as holds if $q_k$ is a quantile belonging to $\theta_k$).
\end{lemma}
{\bf Proof} of Lemma \ref{lineqdiff}: assume w.l.o.g. $q_1\le q_2,\
0<\theta_1\le\theta_2<1$. Consider
$z_j\le q_1,\ q_1<z_j<q_2,\ q_2\le z_j$ separately;
first $z_j\le q_1$. By definition,
\begin{eqnarray*}
  & |\Phi_j(\z,\theta_1,q_1)-\Phi_j(\z,\theta_2,q_2)| =
|(1-\theta_1)(q_1-z_j)-(1-\theta_2)(q_2-z_j)| &\\
  & = |(q_1-q_2)+(\theta_1+\theta_2)(q_2-q_1)-\theta_1q_2+\theta_2q_1+
z_j(\theta_1-\theta_2)|&\\
  & \le |q_2-q_1|+|\theta_1+\theta_2||q_2-q_1|+\theta_2|q_2-q_1|+
|z_j(\theta_1-\theta_2)|\le  |z_j|
|\theta_2-\theta_1|+4|q_2-q_1|.
\end{eqnarray*}
For $q_1<z_j<q_2$:
\begin{displaymath}
  |\Phi_j(\z,\theta_1,q_1)-\Phi_j(\z,\theta_2,q_2)| =
|\theta_1(z_j-q_1)-(1-\theta_2)(q_2-z_j)| \le |q_2-q_1|.
\end{displaymath}
For $q_2\le z_j$:
\begin{displaymath}
  |\Phi_j(\z,\theta_1,q_1)-\Phi_j(\z,\theta_2,q_2)| =
|\theta_1(z_j-q_1)-\theta_2(z_j-q_2)|
\end{displaymath}
and (\ref{eq:aux}) follows along the lines of the first case.
\begin{lemma}\label{lunifpsi} (\ref{eq:sup}) holds under the conditions of
Theorem \ref{tconsist}.
\end{lemma}
{\bf Proof} of Lemma \ref{lunifpsi}:
Suppose (\ref{eq:sup}) were wrong. This means that
there exist $\epsilon>0,\ \delta>0$, a subsequence $M$ of $(1,2,\ldots)$
and $(\theta^*_m)_{m\in M}$ such that
\begin{equation}
  \label{eq:sub}
  \forall m \in M:\ P\left\{|\Psi_m(\theta^*_m)-\Psi(\theta^*_m)|>
\epsilon\right\}\ge \delta.
\end{equation}
W.l.o.g. (because $(\theta_m)_{m\in M}\in T^M$ is bounded and at
least a subsequence
has a limit) there exists $\theta^*=\lim_{m\to\infty}\theta_m^*$.

Consider
\begin{equation}
  \label{eq:dec2}
  |\Psi_m(\theta^*_m)-\Psi(\theta^*_m)|\le |\Psi_m(\theta^*_m)-\Psi_m(\theta^*)|
+|\Psi_m(\theta^*)-\Psi(\theta^*)|+|\Psi(\theta^*)-\Psi(\theta^*_m)|.
\end{equation}
Continuity of $\Psi$ forces the third term of the right side of
(\ref{eq:dec2}) to converge to 0.

Regarding the second term, define a version of $\Psi_n$ using the true
quantiles instead of the empirical ones:
\begin{eqnarray*}
    \Psi^*_n(\theta) &=& \frac{1}{n}\left(\ds\sum_{i:\ C_i=0}\ind\left[\ds\sum_{j=1}^p (\Phi_{j}(\Z_i,\theta,q_{1j}(\theta))-
\Phi_{j}(\Z_i,\theta,q_{0j}(\theta)))>0\right] + \right.\\
&& \left.\ds\sum_{i:\ C_i=1}\ind\left[\ds\sum_{j=1}^p (\Phi_{j}(\Z_i,\theta,q_{1j}(\theta))-
\Phi_{j}(\Z_i,\theta,q_{1j}(\theta)))\le 0\right]\right).
\end{eqnarray*}
Consider
\begin{displaymath}
  |\Psi_m(\theta^*)-\Psi(\theta^*)|\le |\Psi_m(\theta^*)-\Psi_m^*(\theta^*)|+
|\Psi_m^*(\theta^*)-\Psi(\theta^*)|.
\end{displaymath}
Because of the strong law of large numbers,
$\lim_{m\to\infty}|\Psi_m^*(\theta^*)-\Psi(\theta^*)|=0$ a.s.

For given $\z$ and $\theta$, $\Phi_j$ is continuous in $q$. Furthermore
quantiles are strongly consistent, and therefore (\ref{eq:aux}) will enforce
$\lim_{m\to\infty}|\Psi_m(\theta^*)-\Psi_m^*(\theta^*)|=0$ a.s.

Now consider the first term of the right side of (\ref{eq:dec2}).
\begin{equation}
\label{eq:qs}
 |q_{kjm}(\theta_m^*)-q_{kjm}(\theta^*)|\le |q_{kjm}(\theta^*)-q_{kj}(\theta^*)|
+ |q_{kjm}(\theta_m^*)-q_{kj}(\theta_m^*)|+|q_{kj}(\theta_m^*)-q_{kj}(\theta^*)|.
\end{equation}
From Theorem 3 in \cite{Mas82}, which assumes A1,
$\lim_{m\to\infty} \sup_{\theta\in T}
|q_{kj}(\theta)-q_{kjn}(\theta)|=0$ a.s.. This enforces the first two terms
on the left side of (\ref{eq:qs}) to converge to zero a.s.. The last term
converges to zero because of A1. Therefore
\begin{equation}
\label{eq:qas}
|q_{kjm}(\theta_m^*)-q_{kjm}(\theta^*)|\to 0 \mbox{ a.s.}
\end{equation}
Let $D_{n}(\theta,\z)=\sum_{j=1}^p (\Phi_{1jn}(\z,\theta)-
\Phi_{0jn}(\z,\theta))$, $D(\theta,\z)=\sum_{j=1}^p (\Phi_{1j}(\z,\theta)-
\Phi_{0j}(\z,\theta))$.
For $\epsilon>0$ define
\begin{displaymath}
Z_\epsilon=\left\{\z:\ |D(\theta^*,\z)|>\epsilon\right\}\cap
\left\{\z:\ \sum_{j=1}^p|z_j|\le\frac{1}{\epsilon}\right\},
\end{displaymath}
so that
\begin{eqnarray*}
  & |\Psi_m(\theta^*_m)-\Psi_m(\theta^*)|=\frac{1}{m}\left(
\ds\sum_{i:\ C_i=0,\ \Z_i\not\in Z_\epsilon}
\left[\ind(D_{m}(\theta^*_m,\Z_i)>0)-\ind(D_{m}(\theta^*,\Z_i)>0)\right]+\right. &\\
  & \left.\ds\sum_{i:\ C_i=1,\ \Z_i\not\in Z_\epsilon}
\left[\ind(D_{m}(\theta^*_m,\Z_i)\le 0)-\ind(D_{m}(\theta^*,\Z_i)\le 0)
\right]+\right.
&\\
& \left.\ds\sum_{i:\ C_i=0,\ \Z_i\in Z_\epsilon}
\left[\ind(D_{m}(\theta^*_m,\Z_i)>0)-\ind(D_{m}(\theta^*,\Z_i)>0)\right]+\right. &\\
  & \left.\ds\sum_{i:\ C_i=1,\ \Z_i\in Z_\epsilon}
\left[\ind(D_{m}(\theta^*_m,\Z_i)\le 0)-\ind(D_{m}(\theta^*,\Z_i)\le 0)
\right]\right). &
\end{eqnarray*}
Now for large $m$ and arbitrarily small $\delta>0$,
\begin{eqnarray*}
&  \frac{1}{m}\left|\left(
\ds\sum_{i:\ C_i=0,\ \Z_i\not\in Z_\epsilon}
\left[\ind(D_{m}(\theta^*_m,\Z_i)>0)-\ind(D_{m}(\theta^*,\Z_i)>0)\right]+\right.
\right. &\\
  & \left.\left.\ds\sum_{i:\ C_i=1,\ \Z_i\not\in Z_\epsilon}
\left[\ind(D_{m}(\theta^*_m,\Z_i)\le 0)-\ind(D_{m}(\theta^*,\Z_i)\le 0)
\right]\right)\right|\le 1-P(Z_\epsilon)+\delta \mbox{ a.s.}
\end{eqnarray*}
Furthermore, by (\ref{eq:aux}),
\begin{displaymath}
  |D_{m}(\theta^*_m,\Z_i)-D_{m}(\theta^*,\Z_i)|\le \sum_{j=1}^p
\left(2|Z_j||\theta^*_m-\theta^*|+8|q_{kjm}(\theta_m^*)-q_{kjm}(\theta^*)|
\right).
\end{displaymath}
Because $|\theta^*_m-\theta^*|\to 0$, by (\ref{eq:qas}) and
$\sum_{j=1}^p|Z_j|\le\frac{1}{\epsilon}$ for $\Z\in Z_{\epsilon}$, this
difference becomes arbitrarily small a.s. for large enough $m$, and therefore
for $\Z_i\in Z_\epsilon$, $D_{m}(\theta^*_m,\Z_i)$ and $D_{m}(\theta^*,\Z_i)$
will for large enough $m$ be on the same side of zero and their ``$>0$'' and
``$\le 0$''-indicators will therefore be the same, a.s.

For $\epsilon \searrow 0$, A2 enforces $P(Z_\epsilon)\to 1$. This forces the
first term on the right side of (\ref{eq:dec2}) to zero for large $m$, a.s.,
in contradiction to (\ref{eq:sub}), which in turn proves (\ref{eq:sup}).

\section{A result for $p \to\infty$} \label{sp}
Theorem \ref{tconsist} refers to $n\to\infty$ for fixed finite $p$. In many
modern applications, $p$ is so large and often larger than $n$ that results
for $p\to\infty$ seem more appealing, although such results require
$n\to\infty$ as well and it is not entirely clear whether they give a better
justification of a method for applications with given $n$ and $p$. In any case
they contribute to the exploration of a classifier's properties.

\cite{HTX09} prove under some conditions that the misclassification
probability of the median classifier converges to zero for $n, p \to\infty$.
Unfortunately we were not able to prove a result ensuring that
the quantile classifier is, asymptotically,
always at least as good and sometimes better
than the median classifier, as one would hope. Analyzing the proof in
\cite{HTX09}, it can be seen that it adapts in a more or less
straightforward manner to classifiers based on any fixed quantile
other than the median. Despite the fact that one may expect the quantile
classifier to do at least
as good a job (because it incorporates finding the optimal quantile),
this classifier is more difficult to handle theoretically.

We present a result that requires stronger assumptions than those in
\cite{HTX09}, namely considering them uniformly for a range of
quantiles. The arguments in \cite{HTX09} then ensure that the
zero misclassification result carries over to classifiers based on whatever
quantile selection rule is chosen, obviously including selecting the
empirically optimal one.
We restrict ourselves to applying this idea to Theorem
1 in \cite{HTX09}.

Let again $T=[\tau,1-\tau]$ for arbitrarily small
$0<\tau<\frac{1}{2}$.
Let ${\bf U}=(U_1,U_2,\ldots)$ denote an infinite sequence of random variables,
each $U_i$ with uniquely defined $\theta$-quantiles $q_{i}(\theta)$
for all $\theta\in T$ and
median zero. For infinite sequences of constants $(\nu_{X1,\frac{1}{2}},
\nu_{X2,\frac{1}{2}},\ldots),\ (\nu_{Y1,\frac{1}{2}},
\nu_{Y2,\frac{1}{2}},\ldots)$, assume that for each $p$, the $p$-vectors
${\bf X}_1,\ldots,{\bf X}_m$ are identically distributed as
$(\nu_{X1,\frac{1}{2}}+U_1,\ldots,\nu_{Xp,\frac{1}{2}}+U_p)$, and the $p$-vectors
${\bf Y}_1,\ldots,{\bf Y}_n$ are identically distributed as
$(\nu_{Y1,\frac{1}{2}}+U_1,\ldots,\nu_{Yp,\frac{1}{2}}+U_p)$. Define for $i\ge 1$
the quantiles $\nu_{Xi,\theta}=\nu_{Xi,\frac{1}{2}}+q_i(\theta),\
\nu_{Yi,\theta}=\nu_{Yi,\frac{1}{2}}+q_i(\theta)$. Let $C$ be a $[0,1]$-valued RV and
assume {\bf Z} to be distributed
as ${\bf X}_1$ if $C=0$ and as ${\bf Y}_1$ if $C=1$, and
${\bf X}_1,\ldots,{\bf X}_m$,
${\bf Y}_1,\ldots,{\bf Y}_n$ and $({\bf Z},C)$ as totally independent.

Assumptions:
\begin{description}
\item[B1] $\lim_{\lambda\to\infty}\sup_{k\ge 1} E\{|U_k|\ind(|U_k|>\lambda)\}=0.$
\item[B2] For each $c>0:$
$$
\inf_{k\ge 1} \inf_{|x|\ge c} \inf_{\theta\in T}
\left[E\Phi_k({\bf U},\theta,q_k(\theta)+x)-E\Phi_k({\bf U},\theta,q_k(\theta))
\right]>0.
$$
\item[B3] For each $\epsilon>0:$
$$
\inf_{k\ge 1} \inf_{\theta\in T}\left[\min\{\theta-P[U_k\le q_k(\theta)-\epsilon],
\theta-P[U_k\ge q_k(\theta)+\epsilon]\}\right]>0.
$$
\item[B4] With ${\cal B}$ denoting the class of Borel subsets of the real line,
\begin{displaymath}
\lim_{k\to\infty} \sup_{k_1, k_2:\ |k_1-k_2|\ge k}\sup_{B_1,B_2\in{\cal B}}
\left|P(U_{k_1}\in B_1,U_{k_2}\in B_2)-P(U_{k_1}\in B_1)P(U_{k_2}\in B_2)
\right|=0.
\end{displaymath}
\item[B5] The differences $|\nu_{Xk,\theta}-\nu_{Yk,\theta}|$ are uniformly bounded.
\item[B6] For sufficiently small $\epsilon>0$, the proportion of values
$k\in [1,p]$ for which $|\nu_{Xk,\theta}-\nu_{Yk,\theta}|>\epsilon\ \forall
\theta\in T$ is bounded away from zero as $p$ diverges.
\end{description}
The assumptions B1 and B4 are identical to (4.1) and (4.4) in
\cite{HTX09}. B2, B3, B5 and B6 are (4.2), (4.3), (4.5), (4.6)
in \cite{HTX09} enforced to hold uniformly for all $\theta\in T$.
B4 and
B6 enforce a steady flow of relevant information to be added by the data
for increasing $p$. Note that both conditions together mean that at any stage
an infinite amount of relevant information in new variables
independent of what is already known is still waiting to be discovered. This
may look unrealistic but such a thing
is essentially needed for any theory for any method
based on $p\to\infty$ faster than $n$ and $m$. B1 and B5
are needed, given B6, to prevent classification from being dominated by a
single or a finite number of variables, B2 and B3 are about uniform
continuity and well-definedness of the quantiles.
See \cite{HTX09} for further discussion of these assumptions.

Let $R:\ \N \mapsto T$ any quantile selection rule. Let ${\cal R}_{m,n,i},\ i\in\N $
be the sequence of $\{0,1\}$-valued $R(i)$-quantile classifiers computed
from $[({\bf X}_1,0),\ldots,({\bf X}_m,0), ({\bf Y}_1,1),\ldots,({\bf Y}_n,1)]$.
\begin{theorem}\label{tpinfty}
Assume B1-B6 and that both $n$ and $m$ diverge as $p\to\infty$. Then, with
probability converging to 1 as $p$ increases, the classifier ${\cal R}_{m,n,p}$
makes the correct decision, i.e.,
\begin{displaymath}
  P\{{\cal R}_{m,n,p}(\Z)=1|C=0\}+P\{{\cal R}_{m,n,p}(\Z)=0|C=1\} \to 0.
\end{displaymath}
\end{theorem}
{\bf Proof} of Theorem \ref{tpinfty}:
In the proof of Theorem 1 in \cite{HTX09},
B2, B3, B5 and B6 enforce every statement to
hold uniformly for $\theta\in T$, after definitions have been adapted to
general quantile classifiers (i.e., $W_k, D_k, D(Z), S_\lambda, d(Z),
{\cal K}_\epsilon$ and $d_k$  need to be defined as functions of $\theta$ with
quantiles replacing medians, $\Phi_k$ replacing the absolute value where B2
is applied and $q_k(\theta)$ replacing zero where B3 is applied). Equations
(A.1)-(A.6) in  \cite{HTX09} then hold uniformly over $T$.

\begin{remark}
Similar arguments should be possible regarding Theorem 2 in \cite{HTX09}, which
has different assumptions.
\end{remark}

\subsection{Individual treatment of variables}

The empirically optimal quantile classifier as defined above is based on finding
a single $\theta$ that is optimal looking at all variables simultaneously.
One could wonder whether it would be better to choose different $\theta$-values
for each variable. Unfortunately, choosing different $\theta$-values for
different variables is not straightforward. We have tried choosing variable-wise
$\theta$-values by looking at misclassification rates obtained from looking at
$p$ classification problems, each based on a single variable, and then
we used the resulting variable-wise $\theta$-values for a classification rule
incorporating all variables. In most cases this
yielded clearly worse results than selecting a single $\theta$ by looking at all
variables together.

There are two major reasons for this. Firstly, the misclassification rates
based on a single
variable are not very informative for the misclassification result based on all
variables simultaneously. Secondly, using different values of $\theta$ for
different variables results in different scale and distributional shape of the
variable-wise contributions to \eqref{e:distance}, so that certain variables
are implicitly up- and downweighted regardless of their information content for
classification. Using a single optimal $\theta$ for all variables, on the
other hand, gives variables with better discriminative power
some more influence, because they tend to dominate the selection of the optimal
$\theta$, and this is beneficial.

We tried to treat the first problem by defining a one-dimensional
parameter governing convex combinations between the optimal variable-wise
values of $\theta$ and the single optimal value. This parameter was
chosen by optimizing the overall misclassification rate, but on independent
test sets this did not lead to significant improvements compared to the single
optimal $\theta$. There is still some potential for methods finding
individual variable-wise values for $\theta$, but we leave this for further
research.

However, we found a simple method to increase adaptation to the individual
variables, which led to a significant improvement in some
situations while not making things significantly worse elsewhere.

As previously observed in the univariate setting, $\theta$ will depend on the skewness of the involved distributions. In practice, a set of $p>1$ measurements could be skewed in different directions, giving conflicting messages about what values of $\theta$ are to be preferred. In order to overcome this problem, we recommend to change the direction of skewness of variables by applying sign changes in order to unify the direction of skewness.

More specifically, compute a skewness measure separately for each variable, such as the conventional third standardized empirical moment or, alternatively, a measure from the family of the robust quantile-based quantities \citep{Hi75}:
\begin{eqnarray*}
\tau(u)=\frac{F^{-1}(u)+F^{-1}(1-u)-2F^{-1}(1/2)}{F^{-1}(u)-F^{-1}(1-u)},
\end{eqnarray*}
where $F$ denotes the marginal cumulative distribution function and $u$ a fixed value in the interval [0.5,1]. When $u=3/4$ the previous expression corresponds to Galton's measure of skewness, for $u=0.1$ it corresponds to the less robust Kelley's measure of skewness. Evaluate the amount of skewness of each variable separately within classes, in order to avoid overall masking effects due to unbalanced populations, and then summarize by averaging all the within-class measures with equal weights. The signs of variables with negative skewness are then changed, so that finally the variables used for the quantile estimator all have the same (positive) direction of skewness.

This approach takes into account the individuality of the variables in a rather
rough way. Unfortunately in general the connection between skewness and optimal
$\theta$ is not straightforward, so that there is little hope to employ
skewness in a more sophisticated way. The approach recommended here has the
advantage that the choice of $\theta$ is still governed by a one-dimensional
optimization of the overall misclassification rate, and that there is no issue
scaling variable-wise contributions to \eqref{e:distance} against each other.
The results in Sections \ref{scons} and \ref{sp} carry over if the skewness
of all variables is estimated correctly with probability 1 for large enough
$n$.

\section{Numerical results}
\subsection{Simulation study}

We evaluated the performance of the component quantile classifier by a large simulation study comprising several simulated experiments with the aim of assessing the effect of the following factors: sample size, dimensionality, shape of the class-distributions and different level of relevance of the variables for classification.
We generated $p$ vectors from $g=2$ populations in four different main scenarios. In the first scenario we considered symmetric Student's \emph{t}-distributed variables $W_j$ ($j=1,\ldots,p$) with 3 degrees of freedom. We simulated two location-shifted populations from $W_j$ as $X_j=W_j$ and $Y_j=W_j+0.5$. In the second setting we tested the behavior of the classifiers in highly skewed data, by generating identically distributed vectors, $W_j$ with $j=1,\ldots,p$, from a multivariate Gaussian distribution, and transforming them using the exponential function, $X_j=exp(W_j)$ and $Y_j=exp(W_j)+0.2$.
In the third scenario we considered differing distributions for the $p$ variables. More specifically, we first generated $W_j$ from a multivariate Gaussian distribution and then we split $p$ in 5 balanced blocks of different transformations:
\begin{enumerate}
    \item $X_j=W_j$ and $Y_j=W_j+0.2,$
    \item $X_j=exp(W_j)$ and $Y_j=exp(W_j)+0.2,$
    \item $X_j=log(|W_j|)$ and $Y_j=log(|W_j|)+0.2,$
    \item $X_j=W_j^2$ and $Y_j=W_j^2+0.2,$
    \item $X_j=\sqrt{|W_j|}$ and $Y_j=\sqrt{|W_j|}+0.2.$
    \end{enumerate}
In the fourth scenario we simulated differing distributional shapes and levels of skewness even for different classes within the same variable. Here, within each class, data was generated according to Beta distributions with parameters $a$ and $b$ in the interval (0.1,10) randomly generated for each class within each variable. Within each class data have been centered about 0, so that information about class differences is only in the distributional shape, not in the location.

For each of the four scenarios we evaluated the combination of several factors: $p=50,100,500$, $n=50,100,500$, different percentage of relevant variables for classification (100\%, 50\%, and 10\%) and independent or dependent variables (the latter except in the fourth scenario), for a total of $189$ different settings. The dependence structure between the variables has been introduced by generating correlated variables $W_j$ ($j=1,\ldots,p$) from a Gaussian distribution with equicorrelated covariance matrix ($\rho=0.2$). The irrelevant 'noise' variables have been generated independently of each other and the relevant variables
by taking the same base distribution as for the informative variables and leaving out the additive constant (in the fourth scenario a new set of parameters was drawn at random for all observations of each noise variable). Variables were standardized to unit within-class pooled variance in the third scenario but not standardized in the three others, because in the third scenario the scales of the variables seem incompatible, whereas in reality for datasets like those
from the other scenarios the
reasons against standardization given in Remark \ref{rstan} may apply.

For each setting we simulated 100 data sets as training sets and 100 as test sets. The pairs of data sets were split into the two balanced populations with sample size $n/2$.

The component-wise quantile based classifier has been implemented in the R package \verb"quantileDA", (the package will be available on CRAN R homepage soon). Data have been preprocessed according to the skewness correction discussed in Section 2.3 using the conventional skewness measure and the Galton's robust version. In each setting we have evaluated the classifier on a grid of equispaced values $\theta$ in $T=[\tau,1-\tau]$ with $\tau=0.02$. In general, $\tau$ could be tuned to the sample size $n$ as, say, $\tau=5/n$. The optimal $\theta$ has been chosen in each training set. In order to see which $\theta$-values were chosen depending on the model setup, an average value of these has been computed across all the 100 data sets. The mean of the misclassification rates and the standard error of these means were estimated from the classification results in the replicated test sets.

Tables \ref{t:sim1a}-\ref{t:sim4} show the obtained results of the quantile classifiers with data preprocessed according to the Galton and the Skewness corrections (QCG, QCS). The tables show the average misclassification errors and the average of the optimal $\theta$ values across all the 100 data sets in each considered setting. In brackets standard errors have been reported.

We compared the quantile classifier misclassification rates with the ones obtained by nine other classifiers: the component-wise centroid and median classifier
(CC, MC), Fisher's linear discriminant analysis (LDA), the $k$-nearest-neighbor classifier (k-NN; \cite{CH67}), the naive Bayes classifier \citep{HY01}, the support vector machine (SVM; \cite{CV95,WZZ08}), the nearest-shrunken centroid method \citep{THNC02}, penalized logistic regression \citep{PH08} and classification trees (rpart; \cite{BFOS84}). We used the R package \verb"MASS" to implement Fisher's LDA, the library \verb"CLASS" for k-NN with $k=5$, the library \verb"e1071" for the naive Bayes classifier and SVM (Support Vector Machine) with the default settings, the package \verb"pamr" for the nearest-shrunken centroid with threshold set to 1, the package \verb"stepPlr" for penalized logistic regression wit regularization parameter $\lambda=1$, and the package \verb"rpart" for implementing the classification trees.

\begin{figure}[b]
  \includegraphics[scale=0.8]{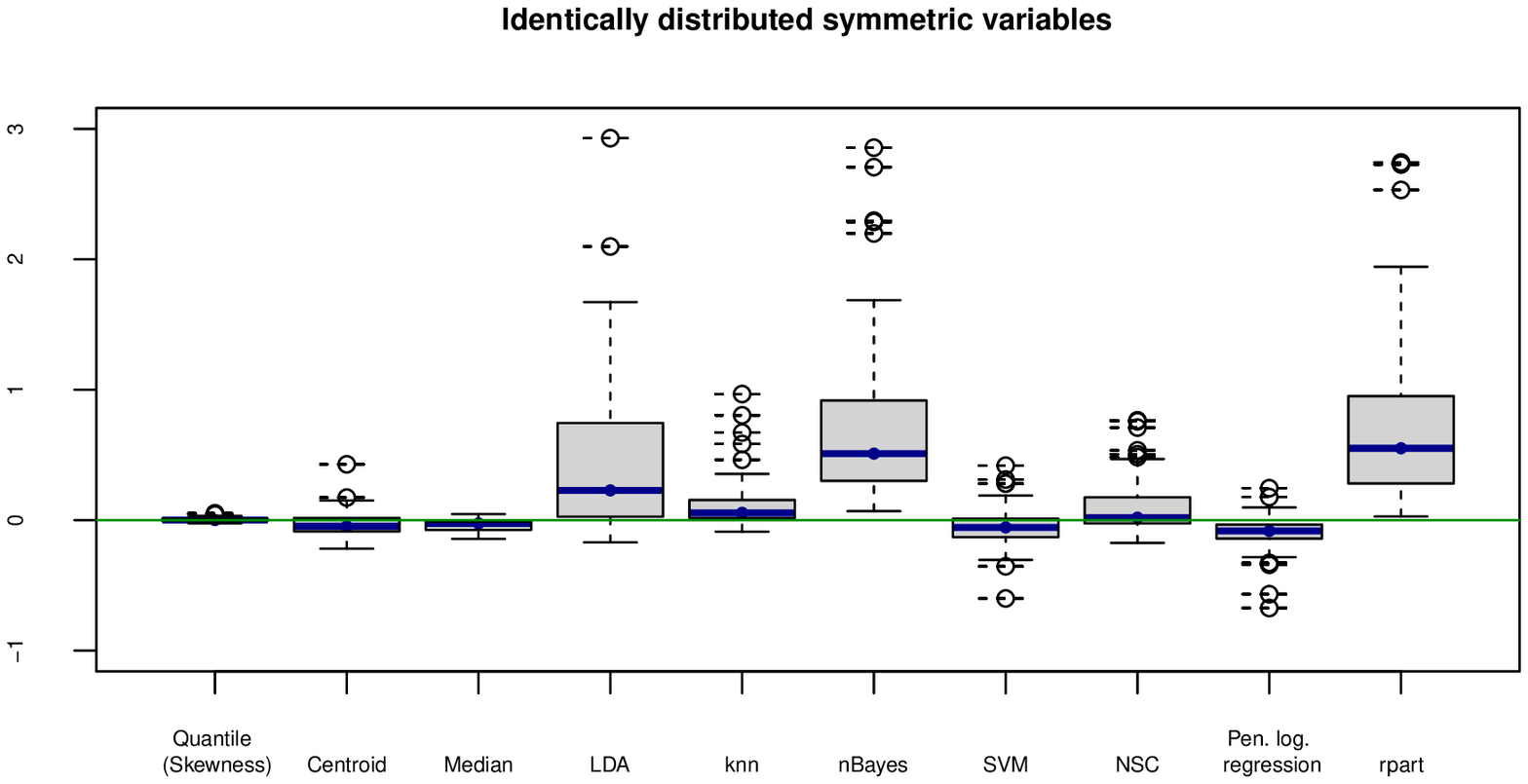}
  \includegraphics[scale=0.8]{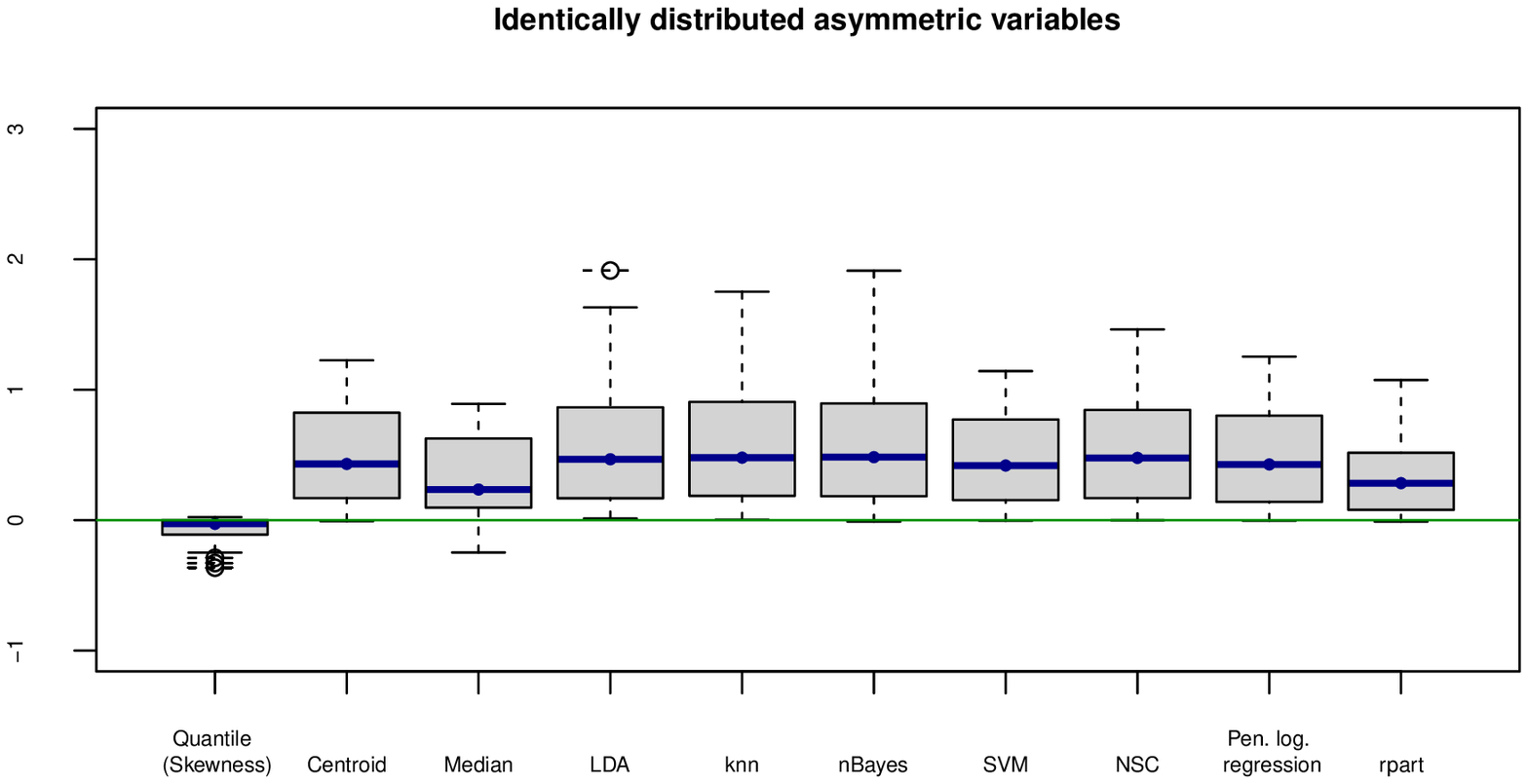}
  \caption{Relative performance of the classifiers with respect to the quantile classifier with Galton skewness correction taken as baseline for all runs in scenarions 1-2.}
  \label{f:boxplots1}
\end{figure}

\begin{figure}[b]
  \includegraphics[scale=0.8]{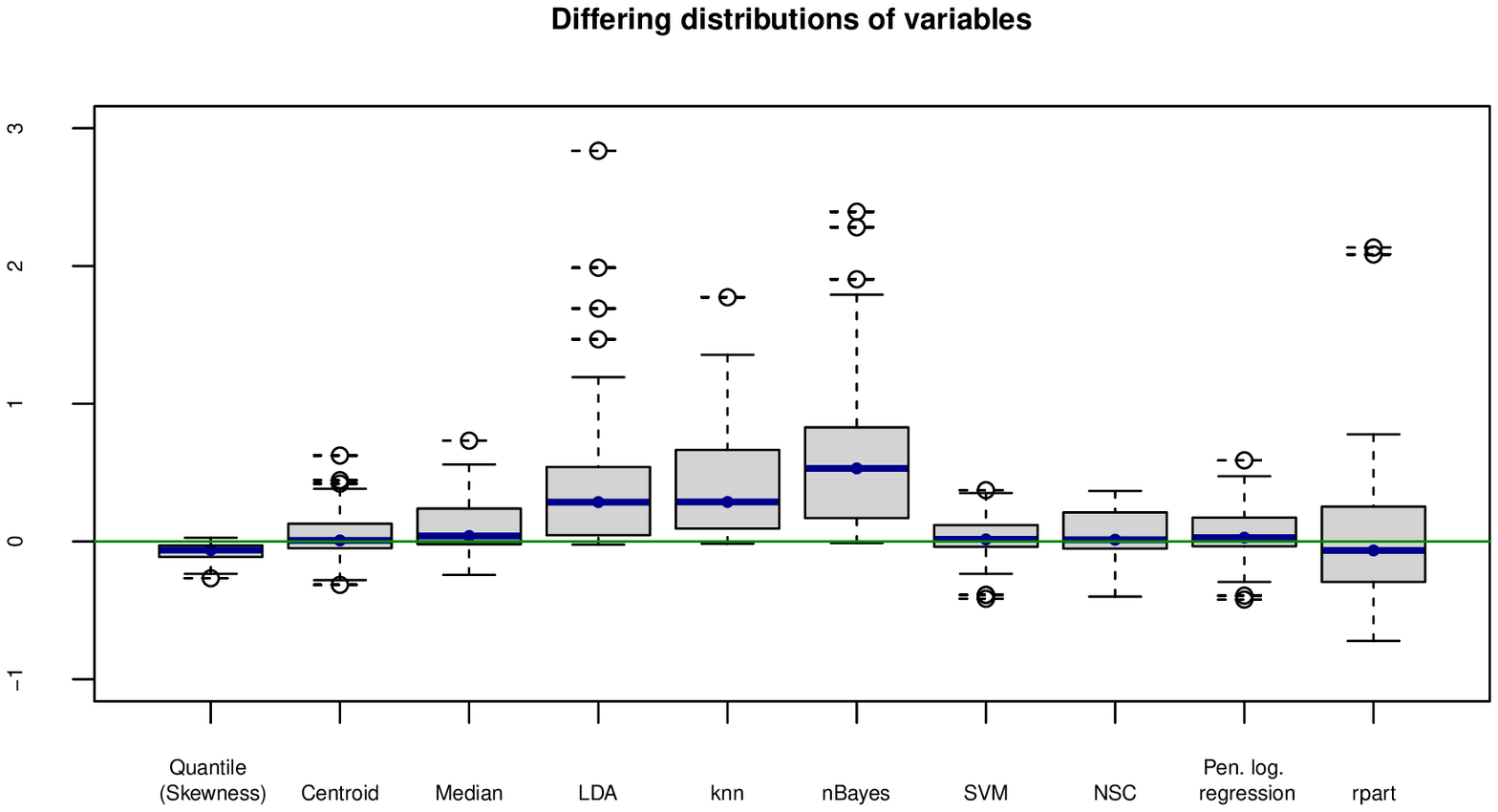}
  \includegraphics[scale=0.8]{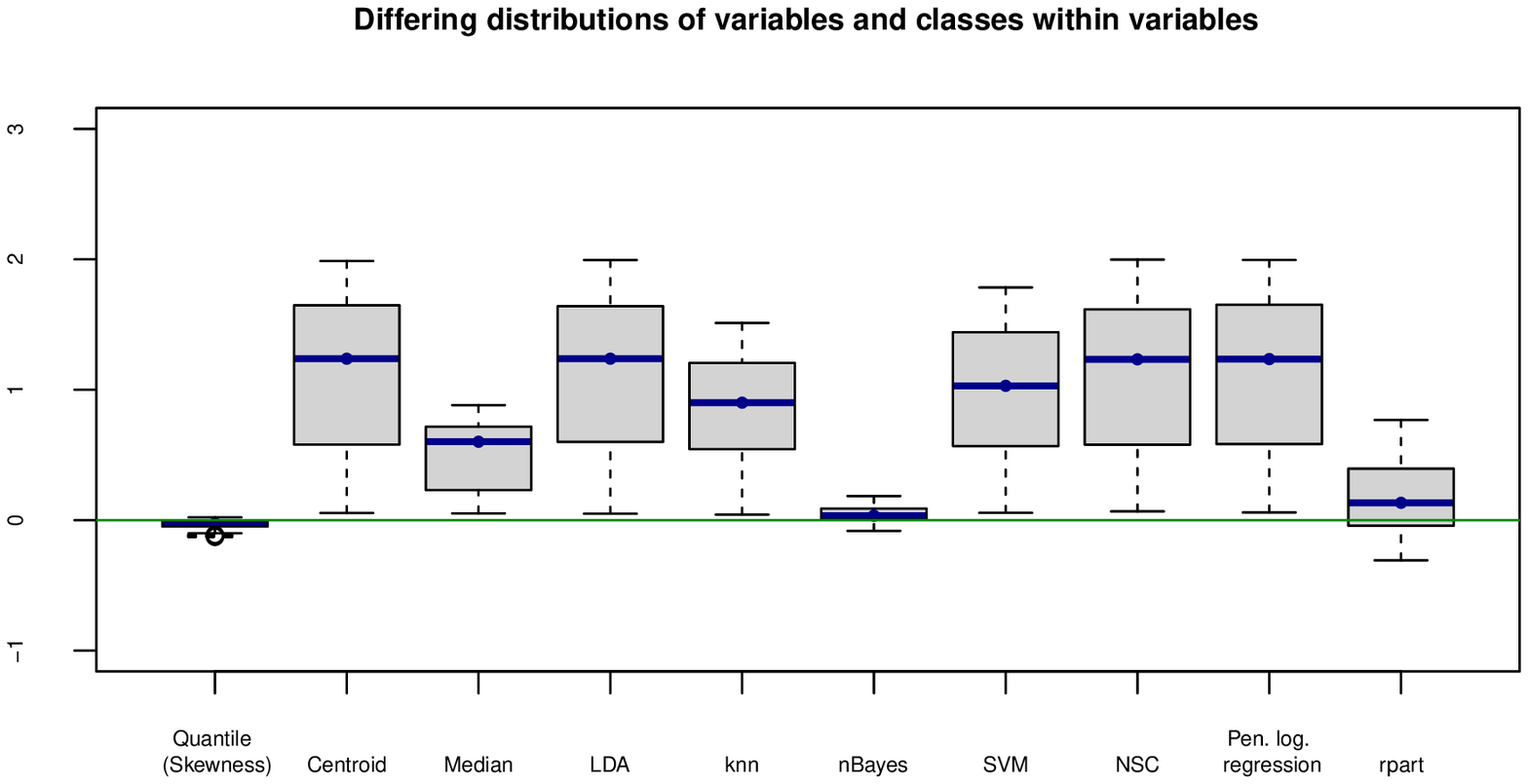}
  \caption{Relative performance of the classifiers with respect to the quantile classifier with Galton skewness correction taken as baseline for all runs in scenarions 3-4.}
  \label{f:boxplots2}
\end{figure}

For all methods, the misclassification rates decrease as the sample size increases. With reference to the quantile classifier the larger the sample size is, the more consistent the choice of the optimal $\theta$ appears and consequently the discriminative power of the method increases. Not surprisingly, the classification performance worsens as the number of irrelevant variables increases.
For fixed sample size and percentage of relevant variables, the methods seem to perform better as $p$ increases, in almost all settings.

To summarize and compare results of the different classifiers, we have computed the relative performance of each classifier with respect to the Galton quantile classifier misclassification rates taken as baseline. More specifically, we have transformed the misclassification rates of each classifier as error rate minus baseline error rate divided by the average error rate in the given setting. The distribution of these rescaled results (aggregated over the different choices of $p, n$, dependence/independence and the percentage of relevant variables) is represented in the boxplots of Figures \ref{f:boxplots1} and \ref{f:boxplots2}.

Results indicate that the quantile classifier performs very well in most situations compared to the other classifiers.
The skewness correction according to the conventional third standardized moment seems to produce a slightly better classification performance in the asymmetric setups. However, the Galton skewness correction is preferable when analyzing real data more sensitive to outliers, as it will be shown in the next section.

In the scenarios with equal distributional shapes and symmetric variables, the performance of the quantile classifiers is similar to the one of centroid and the median classifier, and this is consistent with the chosen optimal value of $\theta$, which is on average close to the midpoint 0.5. stepPlr and SVM perform also very well in this scenario.

In the scenarios with equal distributional shapes and asymmetric variables,
the quantile classifiers
outperform all other methods clearly and more or less uniformly.

With differing distributions of variables, the quantile classifiers again show excellent results. The only method with a better median of the relative performance (see Figures \ref{f:boxplots1} and \ref{f:boxplots2}) is rpart, which is the best method in the setups with few informative variables, due to its use of only a small number of variables. rparts relative performance in the setups with 100\% informative variables and in setups with
small $n$/large$p$ is worse, though.
The overall results of Centroid, SVM, NSC, and
stepPlr are not much worse than those of the quantile classifier, but they are rarely significantly better and sometimes clearly worse.

The fourth scenario with Beta distributions differing between variables and classes within variables is again generally dominated by the quantile classifiers, with nBayes achieving similar results overall and only rpart winning some settings with highest noise ratio. Overall, the methods that compete well with the quantile classifiers in one or two scenarios fall clearly behind in some others.

\begin{table} 
\centering
\scriptsize
\caption{Simulation study: independent identically distributed symmetric variables. Misclassification rates (with standard deviations, i.e. 10*standard errors, in brackets) for different methods. Rows 2 and 4 contain the mean of the chosen values of $\theta$ in the training sets.}
\begin{tabular}{lrrrrrrrrr}
\hline
 & \multicolumn{9}{c}{$n=50$}    \\
  \hline
  & \multicolumn{3}{c}{$p=50$} & \multicolumn{3}{c}{$p=100$} &   \multicolumn{3}{c}{$p=500$}\\
   \hline
 & $100\%$ & $50\%$ & $10\%$ &  $100\%$ & $50\%$ & $10\%$  &  $100\%$ & $50\%$ & $10\%$ \\
 QCG & 0.17 (0.06)& 0.28 (0.06)& 0.42 (0.06)& 0.10 (0.05)& 0.20 (0.07)& 0.41 (0.07)& 0.02 (0.02)& 0.06 (0.04)& 0.32 (0.08)\\
 $\bar{\theta}$ Galton & 0.46 (0.18)& 0.44 (0.18)& 0.44 (0.17)& 0.46 (0.13)& 0.44 (0.14)& 0.43 (0.11)& 0.48 (0.03)& 0.48 (0.02)& 0.48 (0.03)\\
 QCS & 0.17 (0.08)& 0.28 (0.07)& 0.42 (0.06)& 0.10 (0.06)& 0.21 (0.07)& 0.41 (0.06)& 0.02 (0.02)& 0.06 (0.03)& 0.31 (0.07)\\
 $\bar{\theta}$ Skewn.& 0.49 (0.10)& 0.41 (0.18)& 0.43 (0.19)& 0.40 (0.12)& 0.43 (0.15)& 0.44 (0.15)& 0.43 (0.06)& 0.43 (0.03)& 0.44 (0.03)\\
 CC & 0.16 (0.08)& 0.27 (0.07)& 0.43 (0.05)& 0.10 (0.07)& 0.22 (0.10)& 0.42 (0.06)& 0.04 (0.08)& 0.13 (0.14)& 0.37 (0.08)\\
 MC & 0.17 (0.05)& 0.27 (0.06)& 0.42 (0.05)& 0.10 (0.05)& 0.19 (0.06)& 0.42 (0.05)& 0.02 (0.02)& 0.06 (0.04)& 0.32 (0.07)\\
 LDA & 0.38 (0.07)& 0.41 (0.07)& 0.43 (0.05)& 0.23 (0.07)& 0.30 (0.07)& 0.43 (0.05)& 0.26 (0.09)& 0.36 (0.08)& 0.43 (0.05)\\
 knn & 0.19 (0.08)& 0.31 (0.08)& 0.44 (0.05)& 0.14 (0.08)& 0.26 (0.09)& 0.44 (0.05)& 0.08 (0.12)& 0.23 (0.14)& 0.42 (0.07)\\
 n-Bayes & 0.34 (0.11)& 0.42 (0.07)& 0.45 (0.04)& 0.31 (0.14)& 0.40 (0.09)& 0.44 (0.05)& 0.28 (0.15)& 0.36 (0.13)& 0.44 (0.05)\\
 SVM & 0.15 (0.05)& 0.26 (0.07)& 0.42 (0.05)& 0.10 (0.04)& 0.19 (0.07)& 0.42 (0.06)& 0.06 (0.04)& 0.11 (0.07)& 0.42 (0.07)\\
 NSC & 0.29 (0.08)& 0.36 (0.08)& 0.42 (0.06)& 0.23 (0.07)& 0.31 (0.08)& 0.41 (0.06)& 0.10 (0.06)& 0.18 (0.08)& 0.36 (0.08)\\
 stepPlr & 0.14 (0.05)& 0.25 (0.06)& 0.41 (0.05)& 0.07 (0.04)& 0.15 (0.05)& 0.39 (0.06)& 0.01 (0.01)& 0.03 (0.03)& 0.24 (0.06)\\
 rpart & 0.39 (0.05)& 0.41 (0.06)& 0.44 (0.05)& 0.40 (0.06)& 0.39 (0.07)& 0.43 (0.05)& 0.40 (0.06)& 0.41 (0.06)& 0.42 (0.06)\\
   \hline
 & \multicolumn{9}{c}{$n=100$}    \\
  \hline
  & \multicolumn{3}{c}{$p=50$} & \multicolumn{3}{c}{$p=100$} &   \multicolumn{3}{c}{$p=500$}\\
   \hline
 & $100\%$ & $50\%$ & $10\%$ &  $100\%$ & $50\%$ & $10\%$  &  $100\%$ & $50\%$ & $10\%$ \\
  QCG      & 0.15 (0.04)& 0.25 (0.05)& 0.42 (0.04)& 0.09 (0.04)& 0.18 (0.04)& 0.40 (0.05)& 0.01 (0.01)& 0.04 (0.02)& 0.26 (0.04)\\
  $\bar{\theta}$ Galton    & 0.43 (0.15)& 0.43 (0.16)& 0.42 (0.15)& 0.47 (0.18)& 0.44 (0.16)& 0.44 (0.13)& 0.49 (0.05)& 0.48 (0.04)& 0.48 (0.05)\\
  QCS    & 0.16 (0.04)& 0.25 (0.04)& 0.42 (0.04)& 0.10 (0.04)& 0.18 (0.05)& 0.40 (0.05)& 0.01 (0.01)& 0.04 (0.02)& 0.25 (0.04)\\
  $\bar{\theta}$ Skewn.  & 0.44 (0.17)& 0.47 (0.16)& 0.47 (0.19)& 0.42 (0.17)& 0.45 (0.16)& 0.47 (0.14)& 0.46 (0.05)& 0.46 (0.03)& 0.47 (0.05)\\
  CC               & 0.13 (0.06)& 0.22 (0.06)& 0.42 (0.05)& 0.07 (0.03)& 0.16 (0.06)& 0.37 (0.06)& 0.01 (0.01)& 0.04 (0.05)& 0.30 (0.08)\\
  MC                 & 0.14 (0.04)& 0.23 (0.05)& 0.41 (0.05)& 0.08 (0.03)& 0.16 (0.03)& 0.37 (0.05)& 0.01 (0.01)& 0.04 (0.02)& 0.26 (0.05)\\
  LDA                        & 0.18 (0.05)& 0.27 (0.05)& 0.43 (0.04)& 0.35 (0.08)& 0.39 (0.06)& 0.45 (0.04)& 0.12 (0.04)& 0.22 (0.05)& 0.42 (0.05)\\
  knn                        & 0.15 (0.04)& 0.26 (0.06)& 0.43 (0.05)& 0.09 (0.03)& 0.21 (0.07)& 0.42 (0.05)& 0.02 (0.02)& 0.14 (0.10)& 0.41 (0.07)\\
  n-Bayes                    & 0.33 (0.11)& 0.39 (0.09)& 0.46 (0.03)& 0.29 (0.14)& 0.39 (0.09)& 0.45 (0.04)& 0.26 (0.17)& 0.33 (0.15)& 0.45 (0.04)\\
  SVM                        & 0.12 (0.04)& 0.20 (0.04)& 0.40 (0.05)& 0.08 (0.03)& 0.14 (0.03)& 0.37 (0.06)& 0.04 (0.02)& 0.06 (0.03)& 0.32 (0.09)\\
  NSC                        & 0.25 (0.06)& 0.30 (0.05)& 0.41 (0.05)& 0.17 (0.04)& 0.26 (0.06)& 0.38 (0.06)& 0.04 (0.03)& 0.10 (0.04)& 0.28 (0.07)\\
  stepPlr       & 0.13 (0.04)& 0.23 (0.04)& 0.42 (0.05)& 0.06 (0.03)& 0.14 (0.03)& 0.36 (0.05)& 0.01 (0.01)& 0.02 (0.01)& 0.20 (0.04)\\
  rpart                      & 0.38 (0.05)& 0.40 (0.05)& 0.44 (0.04)& 0.39 (0.05)& 0.40 (0.04)& 0.43 (0.05)& 0.38 (0.05)& 0.39 (0.05)& 0.41 (0.05)\\
   \hline
 & \multicolumn{9}{c}{$n=500$}    \\
  \hline
  & \multicolumn{3}{c}{$p=50$} & \multicolumn{3}{c}{$p=100$} &   \multicolumn{3}{c}{$p=500$}\\
   \hline
 & $100\%$ & $50\%$ & $10\%$ &  $100\%$ & $50\%$ & $10\%$  &  $100\%$ & $50\%$ & $10\%$ \\
QCG      & 0.13 (0.02)& 0.20 (0.02)& 0.38 (0.02)& 0.07 (0.01)& 0.13 (0.01)& 0.33 (0.02)& 0.01 (0.01)& 0.03 (0.01)& 0.17 (0.02)\\
$\bar{\theta}$ Galton    & 0.43 (0.11)& 0.43 (0.10)& 0.42 (0.10)& 0.44 (0.13)& 0.44 (0.11)& 0.43 (0.10)& 0.48 (0.20)& 0.45 (0.19)& 0.44 (0.10)\\
QCS    & 0.13 (0.02)& 0.20 (0.02)& 0.37 (0.03)& 0.07 (0.01)& 0.13 (0.01)& 0.32 (0.02)& 0.01 (0.01)& 0.03 (0.01)& 0.17 (0.02)\\
$\bar{\theta}$ Skewn.  & 0.53 (0.12)& 0.50 (0.10)& 0.48 (0.13)& 0.48 (0.13)& 0.50 (0.12)& 0.50 (0.11)& 0.49 (0.17)& 0.51 (0.17)& 0.49 (0.13)\\
CC               & 0.09 (0.01)& 0.17 (0.02)& 0.36 (0.03)& 0.06 (0.04)& 0.10 (0.01)& 0.30 (0.02)& 0.01 (0.00)& 0.02 (0.01)& 0.16 (0.06)\\
MC                 & 0.12 (0.02)& 0.19 (0.02)& 0.36 (0.02)& 0.07 (0.01)& 0.13 (0.02)& 0.32 (0.02)& 0.01 (0.00)& 0.03 (0.01)& 0.16 (0.02)\\
LDA                        & 0.11 (0.02)& 0.19 (0.02)& 0.36 (0.03)& 0.07 (0.01)& 0.13 (0.02)& 0.32 (0.02)& 0.34 (0.06)& 0.38 (0.05)& 0.45 (0.03)\\
knn                        & 0.11 (0.01)& 0.21 (0.02)& 0.41 (0.03)& 0.06 (0.01)& 0.14 (0.02)& 0.38 (0.03)& 0.01 (0.00)& 0.04 (0.02)& 0.33 (0.07)\\
n-Bayes                    & 0.29 (0.12)& 0.36 (0.09)& 0.47 (0.03)& 0.26 (0.15)& 0.37 (0.11)& 0.46 (0.04)& 0.21 (0.17)& 0.35 (0.15)& 0.46 (0.05)\\
SVM                        & 0.10 (0.01)& 0.17 (0.02)& 0.34 (0.02)& 0.06 (0.01)& 0.11 (0.01)& 0.29 (0.02)& 0.03 (0.01)& 0.03 (0.01)& 0.13 (0.02)\\
NSC                        & 0.12 (0.02)& 0.19 (0.02)& 0.33 (0.03)& 0.07 (0.03)& 0.12 (0.02)& 0.28 (0.02)& 0.01 (0.00)& 0.02 (0.01)& 0.13 (0.02)\\
stepPlr       & 0.12 (0.02)& 0.19 (0.02)& 0.36 (0.03)& 0.07 (0.01)& 0.14 (0.02)& 0.32 (0.02)& 0.01 (0.00)& 0.02 (0.01)& 0.14 (0.02)\\
rpart                      & 0.35 (0.02)& 0.36 (0.02)& 0.41 (0.03)& 0.35 (0.02)& 0.36 (0.02)& 0.40 (0.03)& 0.35 (0.02)& 0.36 (0.02)& 0.38 (0.02)\\
   \hline
\end{tabular}\label{t:sim1a}
\end{table} 

\begin{table} 
\centering
\scriptsize
\label{t:sim1b}
\caption{Simulation study: dependent identically distributed symmetric variables. Misclassification rates (with standard deviations, i.e. 10*standard errors, in brackets) for different methods. Rows 2 and 4 contain the mean of the chosen values of $\theta$ in the training sets.}
\begin{tabular}{lrrrrrrrrr}
\hline
 & \multicolumn{9}{c}{$n=50$}    \\
  \hline
  & \multicolumn{3}{c}{$p=50$} & \multicolumn{3}{c}{$p=100$} &   \multicolumn{3}{c}{$p=500$}\\
   \hline
 & $100\%$ & $50\%$ & $10\%$ &  $100\%$ & $50\%$ & $10\%$  &  $100\%$ & $50\%$ & $10\%$ \\
QCG      & 0.27 (0.09)& 0.32 (0.08)& 0.42 (0.06)& 0.24 (0.07)& 0.27 (0.07)& 0.41 (0.06)& 0.21 (0.07)& 0.21 (0.06)& 0.35 (0.08)\\
$\bar{\theta}$ Galton    & 0.37 (0.25)& 0.43 (0.23)& 0.47 (0.18)& 0.39 (0.30)& 0.42 (0.23)& 0.44 (0.16)& 0.36 (0.31)& 0.42 (0.19)& 0.46 (0.09)\\
QCS    & 0.27 (0.08)& 0.31 (0.08)& 0.43 (0.05)& 0.24 (0.08)& 0.27 (0.08)& 0.41 (0.06)& 0.22 (0.06)& 0.22 (0.07)& 0.36 (0.07)\\
$\bar{\theta}$ Skewn.  & 0.34 (0.25)& 0.44 (0.25)& 0.40 (0.19)& 0.30 (0.26)& 0.37 (0.23)& 0.41 (0.14)& 0.22 (0.22)& 0.36 (0.18)& 0.43 (0.09)\\
CC               & 0.24 (0.07)& 0.31 (0.07)& 0.43 (0.05)& 0.21 (0.07)& 0.27 (0.08)& 0.43 (0.06)& 0.19 (0.06)& 0.23 (0.09)& 0.40 (0.08)\\
MC                 & 0.24 (0.06)& 0.29 (0.06)& 0.42 (0.05)& 0.20 (0.06)& 0.25 (0.06)& 0.40 (0.06)& 0.18 (0.05)& 0.21 (0.06)& 0.35 (0.07)\\
LDA                        & 0.43 (0.05)& 0.41 (0.06)& 0.43 (0.05)& 0.32 (0.07)& 0.34 (0.08)& 0.43 (0.05)& 0.22 (0.06)& 0.33 (0.07)& 0.43 (0.05)\\
knn                        & 0.27 (0.06)& 0.33 (0.08)& 0.43 (0.05)& 0.25 (0.07)& 0.31 (0.08)& 0.44 (0.05)& 0.24 (0.08)& 0.30 (0.09)& 0.44 (0.06)\\
n-Bayes                    & 0.35 (0.08)& 0.41 (0.06)& 0.45 (0.04)& 0.34 (0.10)& 0.40 (0.08)& 0.45 (0.04)& 0.34 (0.10)& 0.37 (0.09)& 0.44 (0.05)\\
SVM                        & 0.24 (0.06)& 0.29 (0.07)& 0.42 (0.05)& 0.23 (0.07)& 0.26 (0.07)& 0.42 (0.06)& 0.21 (0.06)& 0.22 (0.07)& 0.41 (0.07)\\
NSC                        & 0.32 (0.07)& 0.37 (0.07)& 0.43 (0.06)& 0.29 (0.07)& 0.33 (0.06)& 0.43 (0.06)& 0.22 (0.06)& 0.25 (0.07)& 0.39 (0.07)\\
stepPlr       & 0.28 (0.07)& 0.29 (0.06)& 0.41 (0.06)& 0.24 (0.07)& 0.23 (0.07)& 0.38 (0.07)& 0.19 (0.05)& 0.12 (0.06)& 0.28 (0.07)\\
rpart                      & 0.39 (0.06)& 0.41 (0.06)& 0.43 (0.05)& 0.39 (0.06)& 0.41 (0.07)& 0.44 (0.05)& 0.40 (0.06)& 0.40 (0.06)& 0.43 (0.05)\\
   \hline
 & \multicolumn{9}{c}{$n=100$}    \\
  \hline
  & \multicolumn{3}{c}{$p=50$} & \multicolumn{3}{c}{$p=100$} &   \multicolumn{3}{c}{$p=500$}\\
   \hline
 & $100\%$ & $50\%$ & $10\%$ &  $100\%$ & $50\%$ & $10\%$  &  $100\%$ & $50\%$ & $10\%$ \\
QCG      & 0.26 (0.05)& 0.30 (0.05)& 0.43 (0.04)& 0.23 (0.06)& 0.26 (0.06)& 0.40 (0.05)& 0.20 (0.05)& 0.21 (0.06)& 0.31 (0.06)\\
$\bar{\theta}$ Galton    & 0.40 (0.24)& 0.41 (0.21)& 0.43 (0.17)& 0.41 (0.30)& 0.42 (0.22)& 0.43 (0.14)& 0.38 (0.33)& 0.36 (0.25)& 0.46 (0.12)\\
QCS    & 0.26 (0.05)& 0.30 (0.06)& 0.43 (0.04)& 0.24 (0.06)& 0.25 (0.06)& 0.40 (0.04)& 0.22 (0.06)& 0.21 (0.05)& 0.30 (0.06)\\
$\bar{\theta}$ Skewn.  & 0.42 (0.24)& 0.47 (0.23)& 0.47 (0.21)& 0.37 (0.29)& 0.42 (0.23)& 0.47 (0.17)& 0.35 (0.34)& 0.38 (0.28)& 0.47 (0.14)\\
CC               & 0.21 (0.04)& 0.27 (0.04)& 0.42 (0.05)& 0.20 (0.04)& 0.24 (0.05)& 0.39 (0.05)& 0.19 (0.06)& 0.20 (0.05)& 0.33 (0.09)\\
MC                 & 0.22 (0.05)& 0.28 (0.04)& 0.42 (0.04)& 0.20 (0.04)& 0.23 (0.04)& 0.38 (0.05)& 0.18 (0.04)& 0.19 (0.04)& 0.30 (0.06)\\
LDA                        & 0.32 (0.05)& 0.31 (0.05)& 0.42 (0.05)& 0.44 (0.05)& 0.42 (0.06)& 0.45 (0.04)& 0.23 (0.04)& 0.26 (0.05)& 0.41 (0.04)\\
knn                        & 0.24 (0.05)& 0.31 (0.06)& 0.44 (0.04)& 0.22 (0.04)& 0.27 (0.05)& 0.43 (0.05)& 0.21 (0.05)& 0.24 (0.07)& 0.42 (0.06)\\
n-Bayes                    & 0.35 (0.09)& 0.41 (0.07)& 0.46 (0.03)& 0.35 (0.10)& 0.40 (0.09)& 0.46 (0.03)& 0.34 (0.09)& 0.36 (0.10)& 0.46 (0.04)\\
SVM                        & 0.23 (0.04)& 0.26 (0.05)& 0.40 (0.05)& 0.21 (0.04)& 0.22 (0.04)& 0.38 (0.05)& 0.20 (0.04)& 0.14 (0.04)& 0.35 (0.07)\\
NSC                        & 0.28 (0.05)& 0.32 (0.06)& 0.42 (0.05)& 0.24 (0.05)& 0.29 (0.06)& 0.40 (0.06)& 0.20 (0.05)& 0.22 (0.04)& 0.30 (0.06)\\
stepPlr       & 0.28 (0.05)& 0.27 (0.05)& 0.41 (0.05)& 0.24 (0.04)& 0.21 (0.04)& 0.36 (0.05)& 0.20 (0.04)& 0.08 (0.03)& 0.21 (0.05)\\
rpart                      & 0.39 (0.05)& 0.40 (0.05)& 0.45 (0.04)& 0.38 (0.05)& 0.40 (0.05)& 0.44 (0.04)& 0.38 (0.04)& 0.40 (0.05)& 0.43 (0.04)\\
   \hline
 & \multicolumn{9}{c}{$n=500$}    \\
  \hline
  & \multicolumn{3}{c}{$p=50$} & \multicolumn{3}{c}{$p=100$} &   \multicolumn{3}{c}{$p=500$}\\
   \hline
 & $100\%$ & $50\%$ & $10\%$ &  $100\%$ & $50\%$ & $10\%$  &  $100\%$ & $50\%$ & $10\%$ \\
QCG      & 0.22 (0.02)& 0.25 (0.02)& 0.38 (0.03)&  0.20 (0.02)& 0.22 (0.02)& 0.34 (0.03)& 0.18 (0.02)& 0.19 (0.02)& 0.24 (0.04)\\
$\bar{\theta}$ Galton    & 0.41 (0.15)& 0.43 (0.11)& 0.44 (0.12)&  0.39 (0.13)& 0.39 (0.13)& 0.47 (0.14)& 0.41 (0.25)& 0.43 (0.25)& 0.45 (0.15)\\
QCS    & 0.22 (0.02)& 0.26 (0.02)& 0.38 (0.02)&  0.20 (0.02)& 0.23 (0.02)& 0.34 (0.03)& 0.18 (0.02)& 0.19 (0.02)& 0.24 (0.04)\\
$\bar{\theta}$ Skewn.  & 0.46 (0.15)& 0.50 (0.13)& 0.49 (0.13)&  0.49 (0.19)& 0.48 (0.18)& 0.51 (0.13)& 0.46 (0.25)& 0.46 (0.28)& 0.49 (0.18)\\
CC               & 0.21 (0.02)& 0.24 (0.02)& 0.37 (0.03)&  0.19 (0.02)& 0.22 (0.03)& 0.33 (0.03)& 0.18 (0.02)& 0.18 (0.02)& 0.23 (0.05)\\
MC                 & 0.22 (0.02)& 0.25 (0.02)& 0.37 (0.02)&  0.20 (0.02)& 0.22 (0.02)& 0.33 (0.03)& 0.18 (0.02)& 0.18 (0.02)& 0.23 (0.04)\\
LDA                        & 0.25 (0.02)& 0.23 (0.02)& 0.36 (0.03)&  0.26 (0.02)& 0.18 (0.02)& 0.32 (0.03)& 0.47 (0.02)& 0.41 (0.04)& 0.45 (0.03)\\
knn                        & 0.23 (0.02)& 0.26 (0.02)& 0.41 (0.03)&  0.21 (0.02)& 0.23 (0.02)& 0.39 (0.03)& 0.19 (0.02)& 0.18 (0.03)& 0.34 (0.05)\\
n-Bayes                    & 0.30 (0.08)& 0.38 (0.07)& 0.47 (0.03)&  0.32 (0.09)& 0.38 (0.09)& 0.46 (0.04)& 0.31 (0.10)& 0.37 (0.10)& 0.47 (0.04)\\
SVM                        & 0.22 (0.02)& 0.21 (0.02)& 0.34 (0.02)&  0.20 (0.02)& 0.16 (0.02)& 0.29 (0.02)& 0.18 (0.02)& 0.06 (0.01)& 0.14 (0.02)\\
NSC                        & 0.22 (0.02)& 0.25 (0.02)& 0.34 (0.03)&  0.20 (0.02)& 0.22 (0.02)& 0.31 (0.03)& 0.18 (0.02)& 0.18 (0.02)& 0.22 (0.02)\\
stepPlr       & 0.25 (0.02)& 0.23 (0.02)& 0.36 (0.03)&  0.26 (0.02)& 0.19 (0.02)& 0.32 (0.03)& 0.23 (0.02)& 0.05 (0.01)& 0.15 (0.02)\\
rpart                      & 0.36 (0.02)& 0.37 (0.02)& 0.42 (0.02)&  0.36 (0.02)& 0.37 (0.02)& 0.40 (0.03)& 0.36 (0.02)& 0.36 (0.02)& 0.39 (0.03)\\
   \hline
\end{tabular}
\end{table} 

\begin{table} 
\centering
\scriptsize
\label{t:sim2a}
\caption{Simulation study: independent identically distributed asymmetric variables. Misclassification rates (with standard deviations, i.e. 10*standard errors, in brackets) for different methods. Rows 2 and 4 contain the mean of the chosen values of $\theta$ in the training sets.}
\begin{tabular}{lrrrrrrrrr}
\hline
 & \multicolumn{9}{c}{$n=50$}    \\
  \hline
  & \multicolumn{3}{c}{$p=50$} & \multicolumn{3}{c}{$p=100$} &   \multicolumn{3}{c}{$p=500$}\\
   \hline
 & $100\%$ & $50\%$ & $10\%$ &  $100\%$ & $50\%$ & $10\%$  &  $100\%$ & $50\%$ & $10\%$ \\
QCG      & 0.25 (0.09)& 0.36 (0.08)& 0.43 (0.05)& 0.26 (0.10)& 0.36 (0.09)& 0.44 (0.05)& 0.26 (0.07)& 0.35 (0.07)& 0.45 (0.04) \\
$\bar{\theta}$ Galton    & 0.18 (0.16)& 0.28 (0.26)& 0.46 (0.31)& 0.35 (0.27)& 0.44 (0.28)& 0.60 (0.26)& 0.48 (0.23)& 0.52 (0.20)& 0.61 (0.11) \\
QCS    & 0.20 (0.07)& 0.28 (0.08)& 0.42 (0.05)& 0.21 (0.08)& 0.24 (0.07)& 0.42 (0.06)& 0.27 (0.10)& 0.26 (0.07)& 0.30 (0.09) \\
$\bar{\theta}$ Skewn.  & 0.06 (0.05)& 0.08 (0.10)& 0.29 (0.30)& 0.06 (0.10)& 0.10 (0.18)& 0.38 (0.38)& 0.06 (0.10)& 0.05 (0.09)& 0.15 (0.30) \\
CC               & 0.43 (0.05)& 0.44 (0.05)& 0.46 (0.04)& 0.43 (0.05)& 0.43 (0.05)& 0.44 (0.05)& 0.39 (0.06)& 0.43 (0.05)& 0.45 (0.04) \\
MC                 & 0.38 (0.07)& 0.43 (0.05)& 0.44 (0.05)& 0.34 (0.07)& 0.40 (0.06)& 0.45 (0.04)& 0.17 (0.06)& 0.30 (0.07)& 0.43 (0.05) \\
LDA                        & 0.44 (0.05)& 0.44 (0.04)& 0.44 (0.05)& 0.44 (0.04)& 0.43 (0.04)& 0.45 (0.04)& 0.43 (0.05)& 0.44 (0.05)& 0.45 (0.04) \\
knn                        & 0.45 (0.05)& 0.46 (0.03)& 0.46 (0.04)& 0.44 (0.04)& 0.45 (0.04)& 0.45 (0.04)& 0.45 (0.05)& 0.46 (0.04)& 0.46 (0.03) \\
n-Bayes                    & 0.44 (0.04)& 0.44 (0.05)& 0.44 (0.05)& 0.45 (0.04)& 0.45 (0.04)& 0.45 (0.04)& 0.44 (0.05)& 0.44 (0.05)& 0.44 (0.05) \\
SVM                        & 0.43 (0.04)& 0.44 (0.05)& 0.44 (0.05)& 0.43 (0.05)& 0.43 (0.04)& 0.45 (0.04)& 0.39 (0.06)& 0.43 (0.05)& 0.45 (0.04) \\
NSC                        & 0.45 (0.04)& 0.45 (0.04)& 0.45 (0.04)& 0.45 (0.05)& 0.44 (0.05)& 0.44 (0.04)& 0.43 (0.06)& 0.43 (0.05)& 0.45 (0.04) \\
stepPlr       & 0.43 (0.05)& 0.44 (0.05)& 0.44 (0.04)& 0.44 (0.05)& 0.43 (0.05)& 0.44 (0.05)& 0.38 (0.07)& 0.42 (0.05)& 0.45 (0.04) \\
rpart                      & 0.42 (0.05)& 0.42 (0.06)& 0.44 (0.04)& 0.42 (0.06)& 0.42 (0.06)& 0.44 (0.05)& 0.41 (0.06)& 0.42 (0.06)& 0.44 (0.05) \\
   \hline
 & \multicolumn{9}{c}{$n=100$}    \\
  \hline
  & \multicolumn{3}{c}{$p=50$} & \multicolumn{3}{c}{$p=100$} &   \multicolumn{3}{c}{$p=500$}\\
   \hline
 & $100\%$ & $50\%$ & $10\%$ &  $100\%$ & $50\%$ & $10\%$  &  $100\%$ & $50\%$ & $10\%$ \\
QCG      & 0.09 (0.04)& 0.18 (0.05)& 0.42 (0.05)& 0.07 (0.04)& 0.14 (0.05)& 0.37 (0.06)& 0.05 (0.07)& 0.14 (0.11)& 0.42 (0.06)\\
$\bar{\theta}$ Galton    & 0.04 (0.02)& 0.04 (0.03)& 0.19 (0.25)& 0.05 (0.06)& 0.06 (0.06)& 0.17 (0.23)& 0.27 (0.20)& 0.29 (0.24)& 0.56 (0.25)\\
QCS    & 0.09 (0.03)& 0.17 (0.04)& 0.41 (0.05)& 0.06 (0.03)& 0.12 (0.04)& 0.35 (0.05)& 0.01 (0.02)& 0.06 (0.07)& 0.27 (0.10)\\
$\bar{\theta}$ Skewn.  & 0.04 (0.02)& 0.03 (0.02)& 0.13 (0.21)& 0.03 (0.02)& 0.04 (0.02)& 0.10 (0.17)& 0.17 (0.15)& 0.11 (0.17)& 0.17 (0.31)\\
CC               & 0.43 (0.05)& 0.45 (0.03)& 0.46 (0.03)& 0.41 (0.05)& 0.44 (0.04)& 0.46 (0.03)& 0.33 (0.05)& 0.40 (0.05)& 0.46 (0.03)\\
MC                 & 0.34 (0.06)& 0.41 (0.05)& 0.46 (0.04)& 0.30 (0.04)& 0.38 (0.06)& 0.45 (0.03)& 0.11 (0.03)& 0.25 (0.04)& 0.43 (0.04)\\
LDA                        & 0.44 (0.04)& 0.46 (0.03)& 0.46 (0.03)& 0.46 (0.03)& 0.45 (0.04)& 0.46 (0.03)& 0.41 (0.05)& 0.44 (0.04)& 0.45 (0.03)\\
knn                        & 0.45 (0.03)& 0.46 (0.03)& 0.46 (0.03)& 0.45 (0.03)& 0.46 (0.03)& 0.46 (0.03)& 0.45 (0.04)& 0.46 (0.03)& 0.47 (0.03)\\
n-Bayes                    & 0.46 (0.03)& 0.46 (0.03)& 0.46 (0.03)& 0.46 (0.03)& 0.46 (0.03)& 0.46 (0.03)& 0.46 (0.03)& 0.46 (0.03)& 0.46 (0.03)\\
SVM                        & 0.42 (0.04)& 0.45 (0.04)& 0.46 (0.03)& 0.41 (0.04)& 0.44 (0.05)& 0.46 (0.03)& 0.34 (0.05)& 0.41 (0.05)& 0.46 (0.03)\\
NSC                        & 0.45 (0.03)& 0.46 (0.03)& 0.46 (0.03)& 0.45 (0.04)& 0.46 (0.04)& 0.46 (0.03)& 0.42 (0.05)& 0.44 (0.04)& 0.46 (0.03)\\
stepPlr       & 0.43 (0.04)& 0.46 (0.03)& 0.46 (0.03)& 0.42 (0.05)& 0.45 (0.04)& 0.46 (0.03)& 0.33 (0.05)& 0.40 (0.05)& 0.46 (0.03)\\
rpart                      & 0.36 (0.06)& 0.39 (0.06)& 0.44 (0.05)& 0.37 (0.06)& 0.38 (0.07)& 0.43 (0.05)& 0.37 (0.06)& 0.38 (0.06)& 0.43 (0.05)\\
   \hline
 & \multicolumn{9}{c}{$n=500$}    \\
  \hline
  & \multicolumn{3}{c}{$p=50$} & \multicolumn{3}{c}{$p=100$} &   \multicolumn{3}{c}{$p=500$}\\
   \hline
 & $100\%$ & $50\%$ & $10\%$ &  $100\%$ & $50\%$ & $10\%$  &  $100\%$ & $50\%$ & $10\%$ \\
QCG      & 0.02 (0.01)& 0.09 (0.01)& 0.34 (0.03)& 0.00 (0.00)& 0.03 (0.01)& 0.26 (0.02)& 0.00 (0.00)& 0.00 (0.00)& 0.06 (0.01)\\
$\bar{\theta}$ Galton    & 0.02 (0.01)& 0.02 (0.01)& 0.05 (0.03)& 0.02 (0.01)& 0.02 (0.01)& 0.03 (0.02)& 0.17 (0.04)& 0.02 (0.00)& 0.03 (0.01)\\
QCS    & 0.02 (0.01)& 0.09 (0.01)& 0.34 (0.03)& 0.00 (0.00)& 0.03 (0.01)& 0.26 (0.02)& 0.00 (0.00)& 0.00 (0.00)& 0.06 (0.01)\\
$\bar{\theta}$ Skewn.  & 0.02 (0.01)& 0.02 (0.01)& 0.05 (0.03)& 0.02 (0.01)& 0.02 (0.01)& 0.03 (0.02)& 0.17 (0.04)& 0.02 (0.00)& 0.03 (0.01)\\
CC               & 0.40 (0.02)& 0.44 (0.02)& 0.48 (0.02)& 0.36 (0.02)& 0.42 (0.02)& 0.48 (0.02)& 0.22 (0.02)& 0.33 (0.02)& 0.46 (0.02)\\
MC                 & 0.31 (0.02)& 0.37 (0.02)& 0.46 (0.02)& 0.23 (0.02)& 0.32 (0.02)& 0.45 (0.02)& 0.05 (0.01)& 0.15 (0.02)& 0.38 (0.02)\\
LDA                        & 0.41 (0.02)& 0.44 (0.02)& 0.48 (0.02)& 0.37 (0.02)& 0.42 (0.02)& 0.48 (0.02)& 0.47 (0.02)& 0.48 (0.02)& 0.48 (0.02)\\
knn                        & 0.46 (0.02)& 0.47 (0.02)& 0.48 (0.01)& 0.45 (0.02)& 0.47 (0.02)& 0.48 (0.01)& 0.43 (0.03)& 0.46 (0.02)& 0.48 (0.01)\\
n-Bayes                    & 0.47 (0.02)& 0.48 (0.02)& 0.48 (0.01)& 0.47 (0.02)& 0.48 (0.01)& 0.48 (0.01)& 0.47 (0.02)& 0.48 (0.02)& 0.48 (0.01)\\
SVM                        & 0.37 (0.02)& 0.43 (0.02)& 0.48 (0.02)& 0.33 (0.02)& 0.41 (0.02)& 0.47 (0.02)& 0.19 (0.02)& 0.32 (0.02)& 0.46 (0.02)\\
NSC                        & 0.45 (0.02)& 0.46 (0.02)& 0.48 (0.02)& 0.43 (0.02)& 0.45 (0.02)& 0.48 (0.02)& 0.36 (0.03)& 0.40 (0.02)& 0.46 (0.02)\\
stepPlr       & 0.41 (0.02)& 0.44 (0.02)& 0.48 (0.02)& 0.37 (0.02)& 0.42 (0.03)& 0.48 (0.02)& 0.27 (0.02)& 0.36 (0.02)& 0.47 (0.02)\\
rpart                      & 0.21 (0.03)& 0.22 (0.03)& 0.36 (0.02)& 0.22 (0.03)& 0.21 (0.03)& 0.28 (0.04)& 0.23 (0.03)& 0.23 (0.03)& 0.22 (0.03)\\
   \hline
\end{tabular}
\end{table} 

\begin{table} 
\centering
\scriptsize
\label{t:sim2b}
\caption{Simulation study: dependent identically distributed asymmetric variables. Misclassification rates (with standard deviations, i.e. 10*standard errors, in brackets) for different methods. Rows 2 and 4 contain the mean of the chosen values of $\theta$ in the training sets.}
\begin{tabular}{lrrrrrrrrr}
\hline
 & \multicolumn{9}{c}{$n=50$}    \\
  \hline
  & \multicolumn{3}{c}{$p=50$} & \multicolumn{3}{c}{$p=100$} &   \multicolumn{3}{c}{$p=500$}\\
   \hline
 & $100\%$ & $50\%$ & $10\%$ &  $100\%$ & $50\%$ & $10\%$  &  $100\%$ & $50\%$ & $10\%$ \\
QCG      & 0.36 (0.10)& 0.40 (0.08)& 0.44 (0.04)& 0.37 (0.09)& 0.38 (0.08)& 0.44 (0.04)& 0.39 (0.08)& 0.37 (0.07)& 0.43 (0.04)\\
$\bar{\theta}$ Galton    & 0.30 (0.35)& 0.34 (0.33)& 0.48 (0.28)& 0.44 (0.40)& 0.38 (0.33)& 0.53 (0.34)& 0.53 (0.42)& 0.34 (0.37)& 0.34 (0.35)\\
QCS    & 0.29 (0.11)& 0.33 (0.10)& 0.43 (0.05)& 0.33 (0.11)& 0.32 (0.11)& 0.42 (0.06)& 0.37 (0.12)& 0.31 (0.13)& 0.38 (0.08)\\
$\bar{\theta}$ Skewn.  & 0.18 (0.29)& 0.19 (0.27)& 0.33 (0.28)& 0.34 (0.38)& 0.28 (0.33)& 0.33 (0.31)& 0.54 (0.40)& 0.35 (0.35)& 0.30 (0.35)\\
CC               & 0.45 (0.04)& 0.44 (0.05)& 0.45 (0.04)& 0.44 (0.05)& 0.45 (0.05)& 0.45 (0.05)& 0.44 (0.06)& 0.44 (0.05)& 0.45 (0.05)\\
MC                 & 0.41 (0.06)& 0.43 (0.05)& 0.44 (0.04)& 0.41 (0.06)& 0.43 (0.05)& 0.45 (0.04)& 0.41 (0.06)& 0.41 (0.06)& 0.44 (0.05)\\
LDA                        & 0.45 (0.04)& 0.44 (0.04)& 0.45 (0.04)& 0.44 (0.05)& 0.44 (0.05)& 0.45 (0.04)& 0.44 (0.05)& 0.44 (0.04)& 0.44 (0.04)\\
knn                        & 0.45 (0.04)& 0.45 (0.04)& 0.44 (0.04)& 0.45 (0.04)& 0.45 (0.04)& 0.45 (0.04)& 0.46 (0.04)& 0.46 (0.04)& 0.46 (0.04)\\
n-Bayes                    & 0.44 (0.04)& 0.45 (0.05)& 0.45 (0.04)& 0.44 (0.04)& 0.45 (0.05)& 0.44 (0.05)& 0.44 (0.05)& 0.45 (0.05)& 0.44 (0.04)\\
SVM                        & 0.43 (0.06)& 0.44 (0.05)& 0.44 (0.05)& 0.43 (0.05)& 0.45 (0.04)& 0.44 (0.04)& 0.43 (0.05)& 0.44 (0.04)& 0.44 (0.05)\\
NSC                        & 0.46 (0.03)& 0.45 (0.04)& 0.45 (0.04)& 0.45 (0.04)& 0.45 (0.04)& 0.45 (0.04)& 0.43 (0.06)& 0.44 (0.05)& 0.44 (0.05)\\
stepPlr       & 0.44 (0.05)& 0.44 (0.05)& 0.44 (0.05)& 0.44 (0.05)& 0.44 (0.04)& 0.44 (0.04)& 0.44 (0.05)& 0.42 (0.05)& 0.44 (0.05)\\
rpart                      & 0.43 (0.06)& 0.43 (0.05)& 0.44 (0.05)& 0.43 (0.05)& 0.44 (0.05)& 0.44 (0.04)& 0.43 (0.05)& 0.43 (0.06)& 0.43 (0.05)\\
   \hline
 & \multicolumn{9}{c}{$n=100$}    \\
  \hline
  & \multicolumn{3}{c}{$p=50$} & \multicolumn{3}{c}{$p=100$} &   \multicolumn{3}{c}{$p=500$}\\
   \hline
 & $100\%$ & $50\%$ & $10\%$ &  $100\%$ & $50\%$ & $10\%$  &  $100\%$ & $50\%$ & $10\%$ \\
QCG      & 0.19 (0.05)& 0.25 (0.07)& 0.42 (0.05)& 0.20 (0.09)& 0.25 (0.09)& 0.41 (0.06)& 0.34 (0.14)& 0.32 (0.11)& 0.38 (0.08)\\
$\bar{\theta}$ Galton    & 0.03 (0.06)& 0.05 (0.12)& 0.27 (0.31)& 0.07 (0.18)& 0.10 (0.22)& 0.30 (0.33)& 0.51 (0.42)& 0.42 (0.38)& 0.32 (0.33)\\
QCS    & 0.18 (0.05)& 0.24 (0.06)& 0.42 (0.05)& 0.17 (0.07)& 0.22 (0.08)& 0.40 (0.06)& 0.33 (0.16)& 0.31 (0.14)& 0.34 (0.10)\\
$\bar{\theta}$ Skewn.  & 0.03 (0.06)& 0.05 (0.12)& 0.24 (0.30)& 0.05 (0.15)& 0.09 (0.19)& 0.23 (0.30)& 0.51 (0.42)& 0.45 (0.38)& 0.31 (0.34)\\
CC               & 0.44 (0.04)& 0.45 (0.03)& 0.46 (0.03)& 0.45 (0.04)& 0.45 (0.04)& 0.46 (0.03)& 0.44 (0.04)& 0.45 (0.04)& 0.46 (0.03)\\
MC                 & 0.41 (0.05)& 0.43 (0.05)& 0.46 (0.03)& 0.41 (0.05)& 0.43 (0.05)& 0.46 (0.03)& 0.41 (0.05)& 0.41 (0.06)& 0.45 (0.04)\\
LDA                        & 0.46 (0.03)& 0.45 (0.04)& 0.46 (0.03)& 0.46 (0.03)& 0.46 (0.03)& 0.45 (0.03)& 0.45 (0.03)& 0.46 (0.04)& 0.45 (0.03)\\
knn                        & 0.45 (0.03)& 0.46 (0.03)& 0.46 (0.03)& 0.45 (0.04)& 0.46 (0.04)& 0.46 (0.03)& 0.46 (0.04)& 0.46 (0.04)& 0.46 (0.03)\\
n-Bayes                    & 0.46 (0.03)& 0.46 (0.03)& 0.47 (0.03)& 0.46 (0.03)& 0.46 (0.03)& 0.46 (0.03)& 0.45 (0.03)& 0.46 (0.03)& 0.46 (0.03)\\
SVM                        & 0.44 (0.04)& 0.45 (0.04)& 0.45 (0.04)& 0.43 (0.05)& 0.45 (0.04)& 0.46 (0.03)& 0.42 (0.05)& 0.43 (0.05)& 0.46 (0.03)\\
NSC                        & 0.46 (0.03)& 0.47 (0.03)& 0.47 (0.03)& 0.46 (0.03)& 0.46 (0.03)& 0.46 (0.03)& 0.45 (0.04)& 0.45 (0.04)& 0.47 (0.03)\\
stepPlr       & 0.46 (0.03)& 0.45 (0.04)& 0.46 (0.03)& 0.45 (0.03)& 0.44 (0.04)& 0.46 (0.03)& 0.45 (0.03)& 0.40 (0.05)& 0.45 (0.04)\\
rpart                      & 0.40 (0.05)& 0.42 (0.04)& 0.45 (0.04)& 0.41 (0.05)& 0.42 (0.05)& 0.44 (0.04)& 0.41 (0.05)& 0.41 (0.05)& 0.43 (0.05)\\
   \hline
 & \multicolumn{9}{c}{$n=500$}    \\
  \hline
  & \multicolumn{3}{c}{$p=50$} & \multicolumn{3}{c}{$p=100$} &   \multicolumn{3}{c}{$p=500$}\\
   \hline
 & $100\%$ & $50\%$ & $10\%$ &  $100\%$ & $50\%$ & $10\%$  &  $100\%$ & $50\%$ & $10\%$ \\
QCG      & 0.14 (0.01)& 0.18 (0.02)& 0.36 (0.03)&  0.12 (0.01)& 0.15 (0.02)& 0.29 (0.02)& 0.11 (0.01)& 0.12 (0.05)& 0.19 (0.07)\\
$\bar{\theta}$ Galton    & 0.02 (0.00)& 0.02 (0.00)& 0.06 (0.05)&  0.02 (0.00)& 0.02 (0.00)& 0.04 (0.03)& 0.02 (0.00)& 0.04 (0.12)& 0.10 (0.15)\\
QCS    & 0.14 (0.01)& 0.18 (0.02)& 0.36 (0.03)&  0.12 (0.01)& 0.15 (0.02)& 0.29 (0.02)& 0.11 (0.01)& 0.12 (0.05)& 0.19 (0.07)\\
$\bar{\theta}$ Skewn.  & 0.02 (0.00)& 0.02 (0.00)& 0.06 (0.05)&  0.02 (0.00)& 0.02 (0.00)& 0.04 (0.03)& 0.02 (0.00)& 0.04 (0.12)& 0.10 (0.15)\\
CC               & 0.46 (0.02)& 0.46 (0.02)& 0.48 (0.02)&  0.45 (0.02)& 0.45 (0.02)& 0.48 (0.02)& 0.45 (0.02)& 0.44 (0.03)& 0.47 (0.02)\\
MC                 & 0.42 (0.02)& 0.43 (0.02)& 0.47 (0.02)&  0.42 (0.02)& 0.41 (0.02)& 0.46 (0.02)& 0.41 (0.02)& 0.40 (0.04)& 0.44 (0.04)\\
LDA                        & 0.47 (0.02)& 0.45 (0.02)& 0.48 (0.01)&  0.48 (0.02)& 0.43 (0.02)& 0.47 (0.02)& 0.48 (0.01)& 0.47 (0.02)& 0.48 (0.01)\\
knn                        & 0.46 (0.02)& 0.46 (0.02)& 0.48 (0.02)&  0.47 (0.02)& 0.47 (0.02)& 0.48 (0.01)& 0.48 (0.01)& 0.47 (0.02)& 0.48 (0.02)\\
n-Bayes                    & 0.47 (0.02)& 0.48 (0.01)& 0.48 (0.01)&  0.47 (0.02)& 0.48 (0.02)& 0.48 (0.01)& 0.47 (0.02)& 0.47 (0.02)& 0.48 (0.01)\\
SVM                        & 0.44 (0.02)& 0.43 (0.02)& 0.47 (0.02)&  0.43 (0.02)& 0.42 (0.02)& 0.47 (0.02)& 0.41 (0.02)& 0.36 (0.02)& 0.45 (0.02)\\
NSC                        & 0.46 (0.02)& 0.47 (0.02)& 0.48 (0.02)&  0.46 (0.02)& 0.46 (0.02)& 0.48 (0.02)& 0.45 (0.02)& 0.45 (0.02)& 0.47 (0.02)\\
stepPlr       & 0.47 (0.02)& 0.45 (0.02)& 0.47 (0.02)&  0.48 (0.02)& 0.43 (0.02)& 0.47 (0.02)& 0.48 (0.01)& 0.35 (0.02)& 0.44 (0.02)\\
rpart                      & 0.29 (0.03)& 0.31 (0.03)& 0.38 (0.02)&  0.29 (0.03)& 0.31 (0.03)& 0.36 (0.03)& 0.29 (0.03)& 0.31 (0.03)& 0.33 (0.03)\\
   \hline
\end{tabular}
\end{table} 

\begin{table} 
\centering
\scriptsize
\label{t:sim3a}
\caption{Simulation study: independent not identically distributed variables. Misclassification rates (with standard deviations, i.e. 10*standard errors in brackets) for different methods. Rows 2 and 4 contain the mean of the chosen values of $\theta$ in the training sets.}
\begin{tabular}{lrrrrrrrrr}
\hline
 & \multicolumn{9}{c}{$n=50$}    \\
  \hline
  & \multicolumn{3}{c}{$p=50$} & \multicolumn{3}{c}{$p=100$} &   \multicolumn{3}{c}{$p=500$}\\
   \hline
 & $100\%$ & $50\%$ & $10\%$ &  $100\%$ & $50\%$ & $10\%$  &  $100\%$ & $50\%$ & $10\%$ \\
QCG                   & 0.25 (0.08)& 0.36 (0.07)& 0.43 (0.05)& 0.19 (0.09)& 0.33 (0.08)& 0.43 (0.05)& 0.06 (0.04)& 0.17 (0.06)& 0.43 (0.05)\\
$\bar{\theta}$ Galton & 0.27 (0.28)& 0.41 (0.32)& 0.55 (0.32)& 0.25 (0.26)& 0.45 (0.34)& 0.58 (0.30)& 0.02 (0.04)& 0.03 (0.07)& 0.33 (0.18)\\
QCS	                  & 0.22 (0.07)& 0.33 (0.08)& 0.43 (0.05)& 0.15 (0.08)& 0.27 (0.08)& 0.44 (0.05)& 0.03 (0.03)& 0.12 (0.05)& 0.43 (0.06)\\
$\bar{\theta}$ Skewn. & 0.15 (0.20)& 0.23 (0.28)& 0.53 (0.32)& 0.17 (0.23)& 0.25 (0.26)& 0.47 (0.35)& 0.02 (0.04)& 0.03 (0.06)& 0.32 (0.13)\\
CC                    & 0.24 (0.07)& 0.36 (0.08)& 0.44 (0.05)& 0.17 (0.05)& 0.29 (0.07)& 0.43 (0.05)& 0.02 (0.02)& 0.13 (0.05)& 0.44 (0.05)\\
MC                    & 0.26 (0.06)& 0.36 (0.07)& 0.44 (0.05)& 0.19 (0.07)& 0.32 (0.07)& 0.43 (0.05)& 0.03 (0.02)& 0.14 (0.05)& 0.44 (0.05)\\
LDA	                  & 0.41 (0.07)& 0.43 (0.05)& 0.45 (0.05)& 0.32 (0.07)& 0.38 (0.06)& 0.44 (0.04)& 0.25 (0.07)& 0.36 (0.07)& 0.45 (0.04)\\
knn	                  & 0.34 (0.06)& 0.41 (0.06)& 0.44 (0.05)& 0.30 (0.07)& 0.38 (0.06)& 0.44 (0.05)& 0.13 (0.06)& 0.28 (0.07)& 0.44 (0.05)\\
n-Bayes               & 0.40 (0.06)& 0.43 (0.05)& 0.45 (0.04)& 0.37 (0.07)& 0.43 (0.05)& 0.45 (0.04)& 0.34 (0.07)& 0.41 (0.05)& 0.45 (0.05)\\
SVM                   & 0.28 (0.06)& 0.36 (0.07)& 0.43 (0.05)& 0.20 (0.06)& 0.32 (0.07)& 0.43 (0.05)& 0.03 (0.03)& 0.14 (0.06)& 0.50 (0.01)\\
NSC                   & 0.32 (0.07)& 0.37 (0.07)& 0.43 (0.06)& 0.24 (0.07)& 0.33 (0.07)& 0.42 (0.06)& 0.07 (0.04)& 0.16 (0.05)& 0.44 (0.04)\\
stepPlr               & 0.28 (0.07)& 0.36 (0.08)& 0.43 (0.05)& 0.19 (0.05)& 0.32 (0.07)& 0.43 (0.05)& 0.03 (0.03)& 0.13 (0.05)& 0.44 (0.05)\\
rpart                 & 0.33 (0.09)& 0.35 (0.09)& 0.41 (0.07)& 0.32 (0.09)& 0.34 (0.09)& 0.42 (0.06)& 0.31 (0.09)& 0.32 (0.10)& 0.41 (0.06)\\
  \hline
 & \multicolumn{9}{c}{$n=100$}    \\
  \hline
  & \multicolumn{3}{c}{$p=50$} & \multicolumn{3}{c}{$p=100$} &   \multicolumn{3}{c}{$p=500$}\\
   \hline
 & $100\%$ & $50\%$ & $10\%$ &  $100\%$ & $50\%$ & $10\%$  &  $100\%$ & $50\%$ & $10\%$ \\
QCG                   & 0.18 (0.04)& 0.31 (0.07)& 0.44 (0.04)& 0.11 (0.04)& 0.25 (0.05)& 0.45 (0.04)& 0.02 (0.02)& 0.10 (0.09)& 0.40 (0.06)\\
$\bar{\theta}$ Galton & 0.09 (0.10)& 0.22 (0.27)& 0.51 (0.33)& 0.17 (0.14)& 0.22 (0.21)& 0.56 (0.32)& 0.15 (0.18)& 0.32 (0.30)& 0.58 (0.33)\\
QCS                   & 0.17 (0.05)& 0.29 (0.06)& 0.44 (0.04)& 0.11 (0.03)& 0.23 (0.06)& 0.44 (0.04)& 0.01 (0.01)& 0.09 (0.09)& 0.38 (0.06)\\
$\bar{\theta}$ Skewn. & 0.06 (0.10)& 0.11 (0.18)& 0.48 (0.36)& 0.12 (0.12)& 0.18 (0.20)& 0.43 (0.35)& 0.20 (0.18)& 0.25 (0.25)& 0.50 (0.37)\\
CC                    & 0.21 (0.04)& 0.31 (0.06)& 0.44 (0.04)& 0.13 (0.04)& 0.24 (0.05)& 0.43 (0.05)& 0.01 (0.01)& 0.06 (0.03)& 0.35 (0.05)\\
MC                    & 0.24 (0.04)& 0.33 (0.05)& 0.45 (0.04)& 0.16 (0.04)& 0.28 (0.04)& 0.43 (0.04)& 0.01 (0.01)& 0.09 (0.03)& 0.37 (0.05)\\
LDA                   & 0.28 (0.05)& 0.36 (0.05)& 0.45 (0.04)& 0.40 (0.06)& 0.44 (0.05)& 0.46 (0.03)& 0.16 (0.04)& 0.28 (0.06)& 0.44 (0.04)\\
knn                   & 0.32 (0.05)& 0.40 (0.05)& 0.46 (0.03)& 0.27 (0.05)& 0.37 (0.05)& 0.46 (0.03)& 0.10 (0.04)& 0.25 (0.05)& 0.43 (0.04)\\
n-Bayes               & 0.35 (0.05)& 0.42 (0.04)& 0.46 (0.03)& 0.32 (0.05)& 0.40 (0.05)& 0.45 (0.04)& 0.28 (0.05)& 0.37 (0.05)& 0.46 (0.03)\\
SVM                   & 0.24 (0.04)& 0.33 (0.06)& 0.45 (0.04)& 0.15 (0.04)& 0.26 (0.05)& 0.43 (0.04)& 0.01 (0.01)& 0.07 (0.03)& 0.36 (0.05)\\
NSC                   & 0.25 (0.05)& 0.32 (0.06)& 0.42 (0.06)& 0.18 (0.04)& 0.26 (0.05)& 0.39 (0.06)& 0.02 (0.01)& 0.07 (0.03)& 0.29 (0.05)\\
stepPlr               & 0.24 (0.05)& 0.34 (0.05)& 0.44 (0.04)& 0.16 (0.04)& 0.28 (0.05)& 0.44 (0.04)& 0.01 (0.01)& 0.07 (0.03)& 0.36 (0.05)\\
rpart                 & 0.18 (0.06)& 0.21 (0.07)& 0.39 (0.06)& 0.17 (0.06)& 0.19 (0.07)& 0.31 (0.06)& 0.17 (0.05)& 0.17 (0.06)& 0.20 (0.07)\\   \hline
 & \multicolumn{9}{c}{$n=500$}    \\
  \hline
  & \multicolumn{3}{c}{$p=50$} & \multicolumn{3}{c}{$p=100$} &   \multicolumn{3}{c}{$p=500$}\\
   \hline
 & $100\%$ & $50\%$ & $10\%$ &  $100\%$ & $50\%$ & $10\%$  &  $100\%$ & $50\%$ & $10\%$ \\
QCG                    & 0.12 (0.03)& 0.21 (0.04)& 0.40 (0.03)& 0.07 (0.01)& 0.16 (0.01)& 0.37 (0.03)& 0.00 (0.00)& 0.02 (0.01)& 0.50 (0.01)\\
$\bar{\theta}$ Galton  & 0.06 (0.06)& 0.08 (0.08)& 0.22 (0.25)& 0.09 (0.06)& 0.07 (0.06)& 0.22 (0.22)& 0.31 (0.05)& 0.17 (0.09)& 0.23 (0.16)\\
QCS                    & 0.10 (0.02)& 0.17 (0.02)& 0.39 (0.03)& 0.06 (0.01)& 0.13 (0.02)& 0.36 (0.03)& 0.00 (0.00)& 0.01 (0.01)& 0.50 (0.00)\\
$\bar{\theta}$ Skewn.  & 0.02 (0.00)& 0.02 (0.01)& 0.11 (0.16)& 0.05 (0.04)& 0.03 (0.02)& 0.10 (0.16)& 0.33 (0.05)& 0.13 (0.07)& 0.16 (0.15)\\
CC                     & 0.17 (0.02)& 0.26 (0.02)& 0.41 (0.02)& 0.09 (0.01)& 0.18 (0.02)& 0.38 (0.02)& 0.00 (0.00)& 0.02 (0.01)& 0.50 (0.00)\\
MC                     & 0.21 (0.02)& 0.29 (0.02)& 0.42 (0.02)& 0.12 (0.01)& 0.22 (0.02)& 0.40 (0.02)& 0.01 (0.00)& 0.04 (0.01)& 0.50 (0.01)\\
LDA                    & 0.19 (0.02)& 0.27 (0.02)& 0.41 (0.02)& 0.11 (0.01)& 0.20 (0.02)& 0.39 (0.03)& 0.41 (0.05)& 0.43 (0.04)& 0.48 (0.01)\\
knn                    & 0.29 (0.02)& 0.38 (0.02)& 0.47 (0.02)& 0.22 (0.02)& 0.34 (0.03)& 0.46 (0.02)& 0.06 (0.02)& 0.19 (0.02)& 0.49 (0.01)\\
n-Bayes                & 0.27 (0.03)& 0.35 (0.03)& 0.46 (0.02)& 0.22 (0.02)& 0.32 (0.03)& 0.46 (0.02)& 0.11 (0.02)& 0.24 (0.02)& 0.49 (0.01)\\
SVM                    & 0.19 (0.02)& 0.27 (0.02)& 0.42 (0.02)& 0.10 (0.01)& 0.20 (0.02)& 0.38 (0.02)& 0.00 (0.00)& 0.03 (0.01)& 0.50 (0.00)\\
NSC                    & 0.19 (0.02)& 0.27 (0.02)& 0.39 (0.02)& 0.10 (0.01)& 0.18 (0.02)& 0.35 (0.02)& 0.00 (0.00)& 0.02 (0.01)& 0.50 (0.00)\\
stepPlr                & 0.19 (0.02)& 0.27 (0.02)& 0.41 (0.02)& 0.12 (0.02)& 0.21 (0.02)& 0.39 (0.03)& 0.00 (0.00)& 0.03 (0.01)& 0.50 (0.00)\\
rpart                  & 0.04 (0.01)& 0.07 (0.01)& 0.32 (0.03)& 0.04 (0.01)& 0.04 (0.01)& 0.22 (0.03)& 0.05 (0.01)& 0.04 (0.01)& 0.38 (0.03)\\
   \hline
\end{tabular}
\end{table} 

\begin{table} 
\centering
\scriptsize
\caption{Simulation study: dependent not identically distributed variables. Misclassification rates (with standard deviations, i.e. 10*standard errors in brackets) for different methods. Rows 2 and 4 contain the mean of the chosen values of $\theta$ in the training sets.}
\begin{tabular}{lrrrrrrrrr}
\hline
 & \multicolumn{9}{c}{$n=50$}    \\
  \hline
  & \multicolumn{3}{c}{$p=50$} & \multicolumn{3}{c}{$p=100$} &   \multicolumn{3}{c}{$p=500$}\\
   \hline
 & $100\%$ & $50\%$ & $10\%$ &  $100\%$ & $50\%$ & $10\%$  &  $100\%$ & $50\%$ & $10\%$ \\
QCG                    & 0.27 (0.08)& 0.35 (0.08)& 0.43 (0.05)& 0.24 (0.09)& 0.34 (0.09)& 0.44 (0.06)& 0.18 (0.12)& 0.25 (0.11)& 0.41 (0.06)\\
$\bar{\theta}$ Galton  & 0.28 (0.32)& 0.34 (0.34)& 0.53 (0.33)& 0.28 (0.35)& 0.41 (0.37)& 0.61 (0.33)& 0.43 (0.47)& 0.31 (0.43)& 0.34 (0.42)\\
QCS                    & 0.23 (0.07)& 0.33 (0.07)& 0.44 (0.04)& 0.19 (0.09)& 0.30 (0.09)& 0.44 (0.04)& 0.16 (0.15)& 0.22 (0.13)& 0.40 (0.07)\\
$\bar{\theta}$ Skewn.  & 0.13 (0.20)& 0.22 (0.26)& 0.46 (0.35)& 0.18 (0.29)& 0.25 (0.34)& 0.44 (0.39)& 0.34 (0.45)& 0.30 (0.42)& 0.33 (0.43)\\
CC                     & 0.27 (0.06)& 0.34 (0.07)& 0.43 (0.05)& 0.22 (0.06)& 0.32 (0.07)& 0.42 (0.06)& 0.13 (0.07)& 0.23 (0.09)& 0.40 (0.07)\\
MC                     & 0.28 (0.07)& 0.36 (0.06)& 0.44 (0.05)& 0.24 (0.06)& 0.33 (0.07)& 0.42 (0.06)& 0.14 (0.06)& 0.24 (0.08)& 0.40 (0.06)\\
LDA                    & 0.43 (0.05)& 0.43 (0.06)& 0.45 (0.04)& 0.33 (0.07)& 0.39 (0.06)& 0.44 (0.05)& 0.22 (0.06)& 0.35 (0.08)& 0.43 (0.06)\\
knn                    & 0.35 (0.07)& 0.39 (0.06)& 0.44 (0.04)& 0.30 (0.07)& 0.38 (0.06)& 0.44 (0.05)& 0.20 (0.06)& 0.29 (0.07)& 0.43 (0.06)\\
n-Bayes                & 0.38 (0.07)& 0.42 (0.06)& 0.44 (0.04)& 0.36 (0.07)& 0.43 (0.05)& 0.44 (0.05)& 0.32 (0.07)& 0.40 (0.07)& 0.44 (0.05)\\
SVM                    & 0.28 (0.06)& 0.35 (0.06)& 0.44 (0.05)& 0.22 (0.06)& 0.33 (0.07)& 0.43 (0.06)& 0.11 (0.06)& 0.22 (0.08)& 0.41 (0.07)\\
NSC                    & 0.32 (0.08)& 0.36 (0.07)& 0.43 (0.06)& 0.26 (0.06)& 0.33 (0.07)& 0.42 (0.06)& 0.13 (0.05)& 0.20 (0.06)& 0.36 (0.08)\\
stepPlr                & 0.29 (0.07)& 0.36 (0.07)& 0.44 (0.05)& 0.24 (0.06)& 0.32 (0.07)& 0.43 (0.06)& 0.11 (0.04)& 0.17 (0.06)& 0.38 (0.07)\\
rpart                  & 0.33 (0.09)& 0.34 (0.08)& 0.41 (0.07)& 0.34 (0.09)& 0.35 (0.09)& 0.40 (0.07)& 0.32 (0.08)& 0.32 (0.09)& 0.38 (0.08)\\
   \hline
 & \multicolumn{9}{c}{$n=100$}    \\
  \hline
  & \multicolumn{3}{c}{$p=50$} & \multicolumn{3}{c}{$p=100$} &   \multicolumn{3}{c}{$p=500$}\\
   \hline
 & $100\%$ & $50\%$ & $10\%$ &  $100\%$ & $50\%$ & $10\%$  &  $100\%$ & $50\%$ & $10\%$ \\
QCG                   & 0.19 (0.04)& 0.30 (0.06)& 0.45 (0.04)& 0.16 (0.08)& 0.24 (0.06)& 0.44 (0.04)& 0.11 (0.12)& 0.21 (0.13)& 0.41 (0.06)\\
$\bar{\theta}$ Galton & 0.07 (0.09)& 0.22 (0.28)& 0.49 (0.31)& 0.15 (0.25)& 0.13 (0.18)& 0.52 (0.34)& 0.27 (0.42)& 0.33 (0.44)& 0.60 (0.42)\\
QCS                   & 0.18 (0.04)& 0.28 (0.05)& 0.43 (0.05)& 0.14 (0.05)& 0.23 (0.05)& 0.44 (0.04)& 0.09 (0.10)& 0.19 (0.14)& 0.39 (0.07)\\
$\bar{\theta}$ Skewn. & 0.04 (0.05)& 0.10 (0.18)& 0.35 (0.33)& 0.05 (0.10)& 0.07 (0.10)& 0.39 (0.36)& 0.18 (0.35)& 0.28 (0.42)& 0.42 (0.43)\\
CC                    & 0.24 (0.05)& 0.32 (0.05)& 0.43 (0.04)& 0.19 (0.04)& 0.28 (0.05)& 0.44 (0.05)& 0.10 (0.05)& 0.18 (0.07)& 0.39 (0.06)\\
MC                    & 0.27 (0.04)& 0.34 (0.06)& 0.45 (0.04)& 0.21 (0.04)& 0.30 (0.05)& 0.43 (0.04)& 0.11 (0.04)& 0.19 (0.06)& 0.40 (0.05)\\
LDA                   & 0.33 (0.05)& 0.37 (0.05)& 0.44 (0.04)& 0.43 (0.04)& 0.44 (0.04)& 0.46 (0.03)& 0.17 (0.04)& 0.27 (0.05)& 0.42 (0.05)\\
knn                   & 0.34 (0.05)& 0.39 (0.05)& 0.45 (0.04)& 0.28 (0.06)& 0.37 (0.05)& 0.45 (0.04)& 0.15 (0.04)& 0.28 (0.06)& 0.44 (0.04)\\
n-Bayes               & 0.35 (0.05)& 0.41 (0.05)& 0.45 (0.03)& 0.33 (0.06)& 0.40 (0.05)& 0.46 (0.03)& 0.26 (0.05)& 0.37 (0.05)& 0.45 (0.04)\\
SVM                   & 0.26 (0.05)& 0.33 (0.05)& 0.44 (0.04)& 0.19 (0.04)& 0.27 (0.05)& 0.43 (0.04)& 0.07 (0.03)& 0.13 (0.05)& 0.38 (0.05)\\
NSC                   & 0.26 (0.05)& 0.32 (0.06)& 0.42 (0.05)& 0.20 (0.04)& 0.26 (0.05)& 0.39 (0.06)& 0.08 (0.03)& 0.13 (0.03)& 0.29 (0.06)\\
stepPlr               & 0.29 (0.05)& 0.35 (0.05)& 0.44 (0.04)& 0.22 (0.05)& 0.30 (0.05)& 0.44 (0.04)& 0.08 (0.03)& 0.12 (0.03)& 0.35 (0.05)\\
rpart                 & 0.18 (0.06)& 0.22 (0.08)& 0.37 (0.06)& 0.17 (0.06)& 0.20 (0.07)& 0.31 (0.07)& 0.17 (0.05)& 0.17 (0.06)& 0.22 (0.07)\\
  \hline
 & \multicolumn{9}{c}{$n=500$}    \\
  \hline
  & \multicolumn{3}{c}{$p=50$} & \multicolumn{3}{c}{$p=100$} &   \multicolumn{3}{c}{$p=500$}\\
   \hline
 & $100\%$ & $50\%$ & $10\%$ &  $100\%$ & $50\%$ & $10\%$  &  $100\%$ & $50\%$ & $10\%$ \\
QCG                    & 0.14 (0.04)& 0.21 (0.04)& 0.39 (0.04)& 0.10 (0.02)& 0.16 (0.02)& 0.36 (0.03)& 0.02 (0.01)& 0.06 (0.01)& 0.26 (0.04) \\
$\bar{\theta}$ Galton  & 0.06 (0.06)& 0.07 (0.07)& 0.19 (0.22)& 0.06 (0.04)& 0.06 (0.06)& 0.14 (0.18)& 0.03 (0.01)& 0.04 (0.02)& 0.12 (0.15) \\
QCS                    & 0.10 (0.02)& 0.17 (0.03)& 0.38 (0.03)& 0.07 (0.01)& 0.14 (0.02)& 0.36 (0.03)& 0.01 (0.01)& 0.04 (0.01)& 0.24 (0.02) \\
$\bar{\theta}$ Skewn.  & 0.02 (0.00)& 0.03 (0.02)& 0.09 (0.14)& 0.03 (0.01)& 0.02 (0.01)& 0.10 (0.15)& 0.02 (0.01)& 0.03 (0.01)& 0.07 (0.08) \\
CC                     & 0.21 (0.02)& 0.28 (0.02)& 0.41 (0.03)& 0.16 (0.02)& 0.22 (0.02)& 0.38 (0.03)& 0.08 (0.02)& 0.12 (0.02)& 0.29 (0.04) \\
MC                     & 0.24 (0.02)& 0.30 (0.02)& 0.42 (0.03)& 0.18 (0.02)& 0.25 (0.02)& 0.40 (0.03)& 0.09 (0.02)& 0.13 (0.02)& 0.32 (0.04) \\
LDA                    & 0.22 (0.02)& 0.28 (0.02)& 0.41 (0.02)& 0.18 (0.02)& 0.23 (0.02)& 0.39 (0.03)& 0.43 (0.03)& 0.44 (0.03)& 0.47 (0.02) \\
knn                    & 0.30 (0.02)& 0.38 (0.02)& 0.47 (0.02)& 0.24 (0.02)& 0.33 (0.02)& 0.46 (0.02)& 0.13 (0.02)& 0.22 (0.03)& 0.42 (0.02) \\
n-Bayes                & 0.28 (0.03)& 0.36 (0.03)& 0.46 (0.02)& 0.24 (0.03)& 0.33 (0.03)& 0.46 (0.02)& 0.14 (0.03)& 0.24 (0.03)& 0.44 (0.02) \\
SVM                    & 0.21 (0.02)& 0.28 (0.02)& 0.42 (0.02)& 0.14 (0.02)& 0.21 (0.02)& 0.38 (0.03)& 0.03 (0.01)& 0.05 (0.01)& 0.26 (0.02) \\
NSC                    & 0.21 (0.02)& 0.27 (0.02)& 0.39 (0.02)& 0.14 (0.01)& 0.21 (0.02)& 0.35 (0.02)& 0.06 (0.01)& 0.08 (0.01)& 0.21 (0.02) \\
stepPlr                & 0.22 (0.02)& 0.29 (0.02)& 0.41 (0.02)& 0.19 (0.02)& 0.23 (0.02)& 0.39 (0.03)& 0.05 (0.01)& 0.07 (0.01)& 0.28 (0.02) \\
rpart                  & 0.04 (0.01)& 0.07 (0.01)& 0.32 (0.03)& 0.04 (0.01)& 0.04 (0.01)& 0.22 (0.03)& 0.05 (0.01)& 0.05 (0.01)& 0.04 (0.01) \\
   \hline
\end{tabular}\label{t:sim3b}
\end{table} 

\begin{table} 
\centering
\scriptsize
\caption{Simulation study: Beta distributed variables, differing between classes. Misclassification rates (with standard deviations, i.e., 10*standard errors in brackets) for different methods. Rows 2 and 4 contain the mean of the chosen values of $\theta$ in the training sets.}
\begin{tabular}{lrrrrrrrrr}
\hline
 & \multicolumn{9}{c}{$n=50$}    \\
  \hline
  & \multicolumn{3}{c}{$p=50$} & \multicolumn{3}{c}{$p=100$} &   \multicolumn{3}{c}{$p=500$}\\
   \hline
 & $100\%$ & $50\%$ & $10\%$ &  $100\%$ & $50\%$ & $10\%$  &  $100\%$ & $50\%$ & $10\%$ \\
QCG                    & 0.10 (0.07)& 0.24 (0.09)& 0.42 (0.07)& 0.04 (0.06)& 0.14 (0.09)& 0.40 (0.07)& 0.00 (0.00)& 0.00 (0.01)& 0.23 (0.08)\\
$\bar{\theta}$ Galton  & 0.23 (0.37)& 0.26 (0.37)& 0.50 (0.38)& 0.15 (0.31)& 0.19 (0.33)& 0.46 (0.39)& 0.02 (0.00)& 0.02 (0.03)& 0.09 (0.23)\\
QCS                    & 0.07 (0.06)& 0.19 (0.10)& 0.41 (0.07)& 0.03 (0.05)& 0.11 (0.08)& 0.39 (0.08)& 0.00 (0.00)& 0.00 (0.01)& 0.21 (0.09)\\
$\bar{\theta}$ Skewn.  & 0.09 (0.21)& 0.13 (0.25)& 0.31 (0.35)& 0.14 (0.30)& 0.11 (0.25)& 0.35 (0.38)& 0.02 (0.00)& 0.03 (0.10)& 0.13 (0.30)\\
CC                     & 0.45 (0.04)& 0.44 (0.05)& 0.44 (0.04)& 0.45 (0.04)& 0.45 (0.04)& 0.44 (0.04)& 0.44 (0.04)& 0.44 (0.04)& 0.44 (0.05)\\
MC                     & 0.32 (0.09)& 0.38 (0.07)& 0.44 (0.05)& 0.24 (0.06)& 0.35 (0.07)& 0.44 (0.05)& 0.06 (0.03)& 0.20 (0.06)& 0.42 (0.05)\\
LDA                    & 0.44 (0.04)& 0.45 (0.04)& 0.44 (0.04)& 0.45 (0.04)& 0.45 (0.04)& 0.44 (0.04)& 0.45 (0.04)& 0.45 (0.04)& 0.45 (0.04)\\
knn                    & 0.38 (0.06)& 0.42 (0.06)& 0.44 (0.04)& 0.38 (0.07)& 0.43 (0.05)& 0.44 (0.05)& 0.39 (0.06)& 0.43 (0.05)& 0.44 (0.04)\\
n-Bayes                & 0.10 (0.06)& 0.21 (0.09)& 0.39 (0.09)& 0.06 (0.05)& 0.15 (0.07)& 0.38 (0.09)& 0.02 (0.02)& 0.05 (0.03)& 0.29 (0.10)\\
SVM                    & 0.42 (0.05)& 0.43 (0.05)& 0.44 (0.04)& 0.43 (0.05)& 0.44 (0.05)& 0.45 (0.04)& 0.44 (0.04)& 0.44 (0.04)& 0.44 (0.04)\\
NSC                    & 0.44 (0.05)& 0.44 (0.05)& 0.45 (0.04)& 0.44 (0.05)& 0.44 (0.05)& 0.45 (0.04)& 0.45 (0.04)& 0.44 (0.04)& 0.45 (0.04)\\
stepPlr                & 0.45 (0.04)& 0.44 (0.05)& 0.44 (0.04)& 0.44 (0.04)& 0.45 (0.04)& 0.44 (0.04)& 0.44 (0.04)& 0.44 (0.05)& 0.44 (0.05)\\
rpart                  & 0.24 (0.11)& 0.26 (0.11)& 0.35 (0.13)& 0.21 (0.09)& 0.24 (0.12)& 0.32 (0.14)& 0.20 (0.09)& 0.22 (0.10)& 0.26 (0.13)\\
   \hline
 & \multicolumn{9}{c}{$n=100$}    \\
  \hline
  & \multicolumn{3}{c}{$p=50$} & \multicolumn{3}{c}{$p=100$} &   \multicolumn{3}{c}{$p=500$}\\
   \hline
 & $100\%$ & $50\%$ & $10\%$ &  $100\%$ & $50\%$ & $10\%$  &  $100\%$ & $50\%$ & $10\%$ \\
QCG                   & 0.05 (0.04)& 0.17 (0.08)& 0.39 (0.08)& 0.01 (0.02)& 0.08 (0.06)& 0.37 (0.08)& 0.00 (0.00)& 0.01 (0.02)& 0.20 (0.10)\\
$\bar{\theta}$ Galton & 0.05 (0.13)& 0.14 (0.29)& 0.38 (0.40)& 0.06 (0.16)& 0.09 (0.20)& 0.37 (0.40)& 0.05 (0.16)& 0.12 (0.28)& 0.28 (0.42)\\
QCS                   & 0.04 (0.03)& 0.15 (0.07)& 0.38 (0.07)& 0.01 (0.02)& 0.05 (0.03)& 0.34 (0.08)& 0.00 (0.00)& 0.01 (0.03)& 0.19 (0.13)\\
$\bar{\theta}$ Skewn. & 0.03 (0.03)& 0.05 (0.10)& 0.26 (0.33)& 0.03 (0.10)& 0.04 (0.09)& 0.21 (0.33)& 0.10 (0.26)& 0.17 (0.33)& 0.27 (0.39)\\
CC                    & 0.46 (0.03)& 0.46 (0.03)& 0.46 (0.03)& 0.46 (0.03)& 0.46 (0.03)& 0.46 (0.03)& 0.45 (0.03)& 0.46 (0.03)& 0.46 (0.03)\\
MC                    & 0.29 (0.06)& 0.38 (0.06)& 0.45 (0.04)& 0.21 (0.05)& 0.31 (0.06)& 0.44 (0.04)& 0.04 (0.02)& 0.15 (0.04)& 0.41 (0.05)\\
LDA                   & 0.46 (0.03)& 0.46 (0.02)& 0.46 (0.03)& 0.46 (0.04)& 0.46 (0.03)& 0.46 (0.03)& 0.46 (0.03)& 0.46 (0.03)& 0.46 (0.03)\\
knn                   & 0.33 (0.07)& 0.41 (0.06)& 0.45 (0.03)& 0.34 (0.07)& 0.39 (0.07)& 0.45 (0.04)& 0.33 (0.06)& 0.41 (0.05)& 0.46 (0.03)\\
n-Bayes               & 0.07 (0.04)& 0.18 (0.07)& 0.38 (0.08)& 0.04 (0.02)& 0.11 (0.06)& 0.35 (0.09)& 0.01 (0.01)& 0.02 (0.01)& 0.24 (0.07)\\
SVM                   & 0.41 (0.06)& 0.45 (0.05)& 0.46 (0.03)& 0.43 (0.05)& 0.45 (0.03)& 0.46 (0.03)& 0.45 (0.04)& 0.46 (0.03)& 0.46 (0.03)\\
NSC                   & 0.45 (0.04)& 0.46 (0.04)& 0.46 (0.03)& 0.45 (0.04)& 0.46 (0.03)& 0.46 (0.03)& 0.46 (0.03)& 0.46 (0.03)& 0.47 (0.03)\\
stepPlr               & 0.46 (0.03)& 0.46 (0.03)& 0.46 (0.03)& 0.46 (0.03)& 0.46 (0.03)& 0.46 (0.03)& 0.46 (0.03)& 0.46 (0.03)& 0.46 (0.03)\\
rpart                 & 0.15 (0.06)& 0.20 (0.08)& 0.29 (0.10)& 0.14 (0.05)& 0.15 (0.06)& 0.24 (0.10)& 0.12 (0.05)& 0.13 (0.06)& 0.18 (0.07)\\
  \hline
 & \multicolumn{9}{c}{$n=500$}    \\
  \hline
  & \multicolumn{3}{c}{$p=50$} & \multicolumn{3}{c}{$p=100$} &   \multicolumn{3}{c}{$p=500$}\\
   \hline
 & $100\%$ & $50\%$ & $10\%$ &  $100\%$ & $50\%$ & $10\%$  &  $100\%$ & $50\%$ & $10\%$ \\
QCG                    & 0.02 (0.01)& 0.10 (0.04)& 0.35 (0.07)& 0.00 (0.00)& 0.03 (0.02)& 0.27 (0.07)& 0.00 (0.00)& 0.00 (0.00)& 0.05 (0.02)\\
$\bar{\theta}$ Galton  & 0.03 (0.01)& 0.03 (0.02)& 0.20 (0.31)& 0.03 (0.02)& 0.03 (0.02)& 0.07 (0.16)& 0.27 (0.42)& 0.03 (0.09)& 0.03 (0.02)\\
QCS                    & 0.02 (0.01)& 0.10 (0.04)& 0.35 (0.07)& 0.00 (0.00)& 0.03 (0.02)& 0.26 (0.07)& 0.00 (0.00)& 0.00 (0.00)& 0.04 (0.02)\\
$\bar{\theta}$ Skewn.  & 0.03 (0.01)& 0.03 (0.01)& 0.17 (0.29)& 0.02 (0.01)& 0.02 (0.01)& 0.05 (0.05)& 0.32 (0.45)& 0.02 (0.00)& 0.03 (0.02)\\
CC                     & 0.48 (0.01)& 0.48 (0.01)& 0.48 (0.01)& 0.48 (0.01)& 0.48 (0.01)& 0.48 (0.01)& 0.48 (0.01)& 0.48 (0.01)& 0.48 (0.01)\\
MC                     & 0.25 (0.05)& 0.34 (0.05)& 0.44 (0.04)& 0.17 (0.04)& 0.27 (0.05)& 0.42 (0.05)& 0.01 (0.01)& 0.07 (0.02)& 0.33 (0.04)\\
LDA                    & 0.48 (0.01)& 0.48 (0.01)& 0.48 (0.01)& 0.48 (0.01)& 0.48 (0.01)& 0.48 (0.01)& 0.48 (0.01)& 0.48 (0.01)& 0.48 (0.01)\\
knn                    & 0.24 (0.06)& 0.35 (0.05)& 0.46 (0.04)& 0.24 (0.07)& 0.35 (0.06)& 0.46 (0.04)& 0.25 (0.08)& 0.36 (0.04)& 0.46 (0.02)\\
n-Bayes                & 0.05 (0.02)& 0.12 (0.05)& 0.35 (0.08)& 0.02 (0.01)& 0.06 (0.03)& 0.28 (0.09)& 0.00 (0.00)& 0.00 (0.00)& 0.11 (0.04)\\
SVM                    & 0.24 (0.05)& 0.37 (0.05)& 0.47 (0.02)& 0.34 (0.04)& 0.41 (0.03)& 0.47 (0.02)& 0.43 (0.04)& 0.46 (0.03)& 0.48 (0.02)\\
NSC                    & 0.47 (0.04)& 0.47 (0.03)& 0.49 (0.01)& 0.47 (0.03)& 0.48 (0.02)& 0.48 (0.01)& 0.48 (0.01)& 0.48 (0.01)& 0.48 (0.01)\\
stepPlr                & 0.48 (0.01)& 0.48 (0.01)& 0.48 (0.01)& 0.48 (0.01)& 0.48 (0.01)& 0.48 (0.01)& 0.48 (0.01)& 0.48 (0.01)& 0.48 (0.01)\\
rpart                  & 0.06 (0.02)& 0.09 (0.04)& 0.23 (0.09)& 0.05 (0.02)& 0.06 (0.02)& 0.16 (0.07)& 0.03 (0.01)& 0.04 (0.01)& 0.06 (0.02)\\
   \hline
\end{tabular}\label{t:sim4}
\end{table} 

\subsection{Real data example}
\label{sreal}
For illustration, we apply the quantile classifier to a data set from chemistry. These data were collected testing a new method to detect bioaerosol particles based on gaseous plasma electrochemistry. The presence of such particles in air has a big impact on health, but monitoring bioaerosols poses great technical challenges. \cite{SHC12} attempted to tell several different bioaerosols apart based on voltage changes over time on eight different electrodes when particles passed a premixed laminar hydrogen/oxygen/nitrogen flame.

The resulting data are eight time series with 301 observations each for each particle. \cite{SHC12} discussed how the relevant information in every time series can be summarized in six characteristic features, namely
\begin{enumerate}
\item Maximum voltage in series.
\item Minimum voltage in series.
\item Maximum voltage change caused by electrode.
\item Difference between final and initial voltage.
\item Length of positive change caused by the electrode.
\item Length of negative change caused by the electrode.
\end{enumerate}
Details are given in \cite{SHC12}. Actually a seventh variable (time point of
maximum change) was used there, which we omit here. Although in \cite{SHC12} it
contributed to the classification, the chemists (personal communication)
suspected this to be an artifact because knowledge of the experiment
suggests that this variable is caused by other experimental features than
the type of the bioaerosol. We are therefore left with 48 variables (six for each of the eight electrodes).

In the current example, we apply a scheme for variable standardization driven by subject knowledge, which is motivated by the expectation of the chemists that the size of variation in voltage and length of effect is informative and that electrodes and variables for which the electrode causes stronger variation are actually more important for discrimination (low variation often indicates that only noise was picked up by the electrode). Standardization of every variable would remove such information. Still, the variables 1-4 (voltages) on one hand and 6-7 (effect lengths) on the other hand do not have comparable measurement units. Therefore we computed one standard deviation from all $8*4$ voltage variables and standardized all these variables by the same standard deviation, and the $8*2$ effect length variables were also standardized by the standard deviation computed from all of them combined.

We confine ourselves to the classification problem of distinguishing between two bioaerosols, namely Bermuda Smut Spores and Black Walnut Pollen. For each bioaerosol there were data from thirty particles.

The quantile classifier has been applied on no-preprocessed data and on data with signs adjustments according to the conventional skewness and its robust Galton version. We used leave-one-out cross-validation to assess the performance of the classifier. Within each fold we selected the optimal $\theta$ in the training set. Table \ref{t:real} contains the misclassification rates of the quantile classifier according to the different preprocessing strategies. We also evaluated other discriminant methods: the component-wise centroid and median classifiers, linear and quadratic discriminant analysis, the $k$-nearest-neighbor classifier with $k=5$, the naive Bayes classifier, the support vector machine, the nearest-shrunken centroid method, penalized logistic regression and classification trees.

It can be seen from these results that the quantile classifier with Galton skewness correction is particularly effective for classifying the two bioaerosols and outperforms the other methods. Only two particles are misclassified.

It is worth noting that the sign adjustment preprocessing step is particularly relevant. If no sign adjustment is performed, the choice of the optimal quantile value is more variable across the cross validated sets (and closer to the midpoint on average) because of the possible different directions of skewness in the observed variables. In this case, when data are preprocessed according to the Galton skewness correction, the selected optimal $\theta$ across the cross validated sets is always extremely small with an average of 0.04. This means that more discriminant information between the two bioaerosols is contained in the left tail of the observed distributions rather than in their ``core''.

\begin{table}
\centering
\caption{Leave-one-out cross-validated misclassification rates of the bioaerosol particles data. In brackets standard errors are reported.}
\begin{tabular}{lc}
Methods & Misclassification rates \\
\hline
QC (no skewness correction)      	& 0.133 (0.044)\\
QCG      					& 0.033 (0.023)\\
QCS    					& 0.117 (0.042)\\
CC               & 0.217 (0.054) \\
MC                 & 0.267 (0.058) \\
LDA                        & 0.067 (0.032) \\
knn                        & 0.150 (0.046) \\
n-Bayes                    & 0.150 (0.046) \\
SVM                        & 0.100 (0.039) \\
NSC                        & 0.267 (0.058) \\
stepPlr       & 0.100 (0.039) \\
rpart                      & 0.400 (0.064) \\
   \hline
\end{tabular}\label{t:real}
\end{table}

\section{Conclusion}
The idea of the componentwise quantile classifier was inspired by the componentwise median classifier in \cite{HTX09}. The simulations and the application show that the quantile classifier can compete with the median classifier in the (symmetric) situations where the median classifier is best, but is much better for asymmetric and mixed variables due to its larger flexibility. It also compares very favorable to all the other classifiers tested in the present work.

Basic issues with the componentwise quantile classifier are that it ignores the correlation structure (which though does not seem to do much harm in the simulations with dependent variables) and that it requires scaling of the variables because it is not scale equivariant. As all distance-based classifiers, it does not require the classification information to be concentrated on a much lower dimensional space.

First attempts to use different $\theta$-values for different variables
were not successful. This is an issue for future research.

%

\bibliographystyle{chicago}
\bibliography{rif}


\end{document}